\documentclass[fleqn,usenatbib]{mnras}

\usepackage{newtxtext,newtxmath}

\usepackage[T1]{fontenc}

\DeclareRobustCommand{\VAN}[3]{#2}
\let\VANthebibliography\thebibliography
\def\thebibliography{\DeclareRobustCommand{\VAN}[3]{##3}\VANthebibliography}

\usepackage{graphicx}	
\usepackage{amsmath}	

\usepackage{multirow}
\usepackage{diagbox}

\usepackage{CJK}

\newcommand{\citeg}[1]{\citep[e.g.,][]{#1}}

\usepackage{tikz,xcolor,hyperref}

\definecolor{lime}{HTML}{A6CE39}
\DeclareRobustCommand{\orcidicon}{%
	\begin{tikzpicture}
	\draw[lime, fill=lime] (0,0) 
	circle [radius=0.16] 
	node[white] {{\fontfamily{qag}\selectfont \tiny ID}};
	\draw[white, fill=white] (-0.0625,0.095) 
	circle [radius=0.007];
	\end{tikzpicture}
	\hspace{-2mm}
}

\foreach \x in {A, ..., Z}{%
	\expandafter\xdef\csname orcid\x\endcsname{\noexpand\href{https://orcid.org/\csname orcidauthor\x\endcsname}{\noexpand\orcidicon}}
}

\begin{document}
\begin{CJK*}{UTF8}{gbsn}
\title[Intrinsic MS Scatter at $0.5<z<3.0$]{Exploring the Intrinsic Scatter of the Star-Forming Galaxy Main Sequence at redshift 0.5 to 3.0}

\author[R. Huang et al.]{Rongjun Huang (黄钅容钧)\orcidA{}$^{1, 2}$\thanks{E-mail: u6569836@anu.edu.au}, 
Andrew J. Battisti\orcidB{}$^{1, 2}$\thanks{E-mail: andrew.battisti@anu.edu.au}, 
Kathryn Grasha\orcidC{}$^{1, 2, 5}$\thanks{ARC DECRA Fellow}, 
\newauthor
Elisabete da Cunha\orcidD{}$^{2, 3}$, 
Claudia del P Lagos\orcidE{}$^{2, 3}$,  
Sarah K. Leslie\orcidF{}$^{2, 4}$ and 
Emily Wisnioski\orcidG{}$^{1, 2}$
\\  
$^{1}$Research School of Astronomy and Astrophysics, Australian National University, Cotter Road, Weston Creek, ACT 2611, Australia\\ 
$^{2}$ARC Centre of Excellence for All Sky Astrophysics in 3 Dimensions (ASTRO 3D), Australia\\
$^{3}$International Centre for Radio Astronomy Research, University of Western Australia, 35 Stirling Hwy., Crawley, WA 6009, Australia\\
$^{4}$Leiden Observatory, Leiden University, P.O. Box 9513, NL-2300 RA Leiden, The Netherlands\\
$^{5}$Visiting Fellow, Harvard-Smithsonian Center for Astrophysics, 60 Garden Street, Cambridge, MA 02138, USA\\}

\date{Accepted 2023 January 3. Received 2022 December 22; in original form 2022 October 14}

\pubyear{2022}


\label{firstpage}
\pagerange{\pageref{firstpage}--\pageref{lastpage}}
\maketitle
\end{CJK*}

\label{abstract}
\begin{center}
    \begin{abstract}
Previous studies have shown that the normalization and scatter of the galaxy `main sequence' (MS), the relation between star formation rate (SFR) and stellar mass ($M_*$), evolves over cosmic time. However, such studies often rely on photometric redshifts and/or only rest-frame UV to near-IR data, which may underestimate the SFR and $M_*$ uncertainties. We use \texttt{MAGPHYS+photo-z} to fit the UV to radio spectral energy distributions of 12,380 galaxies in the COSMOS field at $0.5<z<3.0$ and self-consistently include photometric redshift uncertainties on the derived SFR and $M_*$. We quantify the effect on the observed MS scatter from (1) photometric redshift uncertainties (which are minor) and (2) fitting only rest-frame ultraviolet to near-infrared observations (which are severe). At fixed redshift and $M_*$, we find that the intrinsic MS scatter for our sample of galaxies is 1.4 to 2.6 times larger than the measurement uncertainty. The average intrinsic MS scatter has decreased by 0.1 dex from $z=0.5$ to $\sim2.0$. At low-$z$, the trend between the intrinsic MS scatter and $M_*$ follows a functional form similar to an inverse stellar mass-halo mass relation (SMHM; $M_*/M_{\rm halo}$ vs $M_*$), with a minimum in intrinsic MS scatter at $\log(M_*/M_{\odot}) \sim10.25$ and larger scatter at both lower and higher $M_*$; while this distribution becomes flatter for high-$z$. The SMHM is thought to be a consequence of feedback effects and this similarity may suggest a link between galaxy feedback and the intrinsic MS scatter. These results favor a slight evolution in the intrinsic MS scatter with both redshift and mass. 
    \end{abstract}
\end{center}





\section{Introduction}
\label{Introduction}

The galaxy main sequence (MS) describes the empirical relation between the star formation rate (SFR) of galaxies and their stellar masses \citeg{daddi07, noeske07, speagle14, whitaker14, renzini15, barro17, leslie20, thorne21}. These studies find that galaxies have higher SFR with increasing redshifts at a fixed stellar mass ($M_*$) in the earlier universe, and more massive galaxies have higher SFRs at a fixed redshift. Some of these studies show a flattening or turnover in the relationship at high masses ($\log(M_*/M_{\odot})>10.5$) \citeg{whitaker14, leslie20, thorne21} and suggest that this turnover is driven by the quenching of star formation due to feedback processes.  

The galaxy MS is a powerful tool for understanding and constraining the distribution and evolution of galaxies \citep{katsianis20, curtis-lake21, popesso22, daddi22}. According to theories of galaxy feedback, the existence of a relatively tight MS is thought to be mainly driven by the dynamical balance between inflows and outflows caused by self-regulated star-formation and/or active galactic nuclei \citep[AGN;][]{somerville&dave15}. Characterising this evolution is difficult because observations only provide a single snapshot in time for each observed galaxy. However, the evolution in the scatter, slope, and normalisation in the MS of large statistical samples of star-forming (SF) galaxies with cosmic time provides an indirect way to study galaxy evolution. The width (or scatter) of the MS at a single redshift is thought to reflect the burstiness of the average star formation history \citeg{guo13, schreiber15, santini17, caplar19, donnari19, katsianis19, Matthee19}. Theories suggest that a small MS width (small scatter, e.g., $\sim0.1$ dex) is indicative of gradual, continuous star formation histories (SFHs). In contrast, large MS widths (large scatter, e.g., $\sim0.4$ dex) are indicative of more bursty, stochastic SFHs \citeg{tacchella16, sparre17}. 

The question around whether the intrinsic MS scatter is constant or evolving is actively debated. Previous studies have found a time-independent MS scatter \citeg{daddi07, noeske07, whitaker12, ciesla14, speagle14, pessa21}, while others suggest it evolves with redshift \citeg{kurczynski16, santini17, katsianis19, tacchella20, davies22, shin22}. As the width of MS is related to the SFH, the MS scatter can provide useful constraints on the evolution of SF galaxies. For example, a larger burstiness or stochasticity in the SFH can lead to an increase in MS scatter, and this may change over cosmic time \citep{Matthee19}.

Improving our understanding of the galaxy MS and its scatter requires using large samples of galaxies with accurate redshifts. However, it is observationally expensive to get spectroscopic redshifts for every galaxy. A common solution is to instead use photometrically derived redshifts ($z_{{\rm phot}}$). Most previous studies of the galaxy MS have relied on determining stellar masses and SFRs based on SED-fitting at fixed photometric redshift $z_{{\rm phot}}$ and ignore the uncertainty of the $z_{{\rm phot}}$ \citeg{speagle14, leslie20, thorne21}. Studies that do not account for zphot uncertainty will systematically underestimate the uncertainties in all distance-dependent parameters (e.g., $M_*$ \& SFR). 

In this study, we use \texttt{MAGPHYS+photo-z} \citep{battisti19} to study the intrinsic scatter of the MS. The improvement in using  \texttt{MAGPHYS+photo-z} is that it sets $z_{{\rm phot}}$ as an unknown quantity and finds its probability distribution \citep{battisti19}. Hence, the uncertainty in the $z_{{\rm phot}}$ is incorporated into the overall uncertainty in the derived physical properties of the galaxy. This allows us to examine how much of the scatter in the MS is driven by measurement uncertainty as opposed to true intrinsic MS scatter or other measurement uncertainties. Simultaneously, \texttt{MAGPHYS+photo-z} also includes IR information to resolve the effect of dust attenuation at UV-near-IR wavelengths on the SED based on dust emission from mid-IR-radio, which dramatically improves the accuracy of the derived properties, particularly for SFRs \citep{battisti19}. Therefore, the unique aspect of \texttt{MAGPHYS+photo-z} is that it uses broadband photometry to predict the best-fitting properties in a self-consistent manner, which helps to mitigate potential biases on the derived values.
 
This paper is organised as follows: Section \ref{Data and Methods} introduces the data and methods used in this study, Section \ref{Results and Analysis} summarises our results, and Section \ref{Discussion} compares our results with some previous observational studies and simulations and Section \ref{Conclusions} outlines our conclusions. Throughout this paper, the flat Lambda Cold Dark Matter ($\Lambda$CDM) model is adopted by assuming the Hubble constant is $H_0$ = 70km/s/Mpc and the mass density of the Universe is $\Omega_{\rm{m,0}}$ = 0.3.

\section{Data and Methods}
\label{Data and Methods}

\subsection{COSMOS Sample}
\label{COSMOS Sample} 

The multi-wavelength observations of galaxies used in this study come from two catalogues: the COSMOS2020 catalogue \citep{weaver22} and the COSMOS Super-deblended catalogue \citep{jin18}. The COSMOS2020 catalogue contains photometric data for $\sim 1$ million sources in 13 filters from UV to near-IR \citep{weaver22}, and the COSMOS Super-deblended catalogue presents photometric data for $\sim 200,000$ galaxies in 11 filters in the mid-IR, far-IR, and radio \citep{jin18}. We cross-match the galaxies' ID and select a subsample of galaxies with SEDs that are sampled well enough to constrain their stellar mass and (dust-corrected) SFRs robustly. To achieve this, we use two criteria: (1) signal-to-noise ratio ($S/N$) of $S/N>3$ in three or more UV--near-IR bands and (2) $S/N$>3 in 2 or more IR--radio bands. By virtue of criterion (2), all of the sources in our sample have a match in the Super-deblended catalogue.

It is important to note that criterion (2) roughly translates into a cut in SFR such that only galaxies above a certain SFR will be included. This SFR threshold increases as a function of increasing redshift (see Section \ref{Influence of IR-selection on SFR- and mass-completeness}). Additionally, AGNs are excluded based on X-ray detections and IR \& radio colour cuts \citep{seymour08, donley12, kirkpatrick13}. This is done because \texttt{MAGPHYS+photo-z} does not include AGN models, so the derived properties are not accurate for these sources. Due to the limited availability of data required for these AGN diagnostics, some AGNs may not be identified and removed. Further details and references on these cuts are in Section 3.3 of the \citet{battisti19}. These selection criteria leave us with a photometric sample of 14,607 galaxies. For later comparison (Section \ref{Accuracy of z_phot relative to z_spec}), only 3,873 of the whole 14,607 galaxies have spectroscopic redshifts ($z_{{\rm spec}}$).

\subsection{Methods}
\label{Methods}

\subsubsection{\texttt{MAGPHYS+photo-z}}
\label{MAGPHYS+photo-z}

\begin{figure*}
    \centering
    \includegraphics[width=0.96\textwidth]{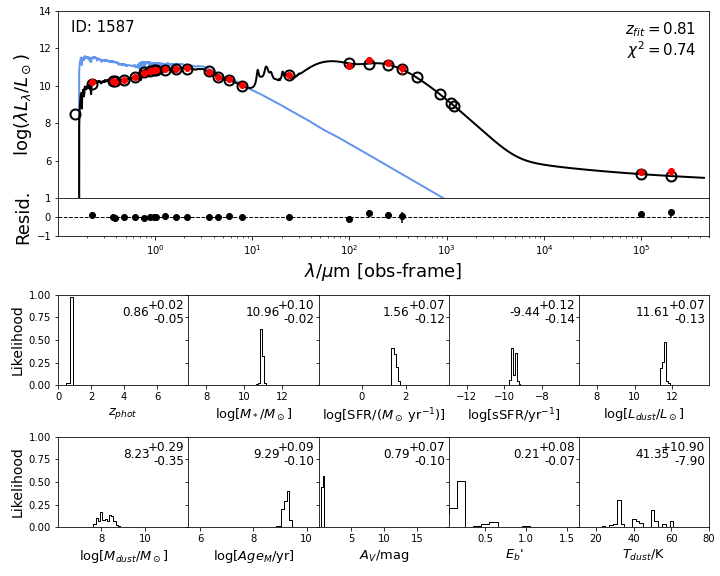}
    \caption{Output of \texttt{MAGPHYS+photo-z} for one of the galaxies of our sample, COSMOS2020 ID 1587. The upper panel shows the best-fit SED (black curve), the observed data (red square) and the predicted unattenuated SED (blue curve). The black open circle on the SED fitting curve is the corresponding model photometry. The goodness of fit is presented by $\chi^2$ in the upper right corner. The lower panel shows the likelihood distribution of 10 basic physical parameters: $z_{{\rm phot}}$, stellar mass ($\log[M_*/M_{\odot}]$), $\log[{\rm SFR}/(M_{\odot}{\rm yr}^{-1})]$, specific-SFR ($\log[{\rm sSFR}/{\rm yr}]$), dust luminosity ($\log[L_{\rm{dust}}/L_{\rm{\odot}}]$), dust mass ($\log[M_{\rm{dust}}/M_{\odot}]$), mass-weighted stellar age ($\log[Age_m$/yr]), V-band dust attenuation ($A_V$/mag), 2175$\overset{\circ}{A}$ bump strength ($E_b'$) and the effective dust temperature ($T_{{dust}}$/K) \citep{battisti19}.}
    \label{fig: MAGPHYS 1587}
\end{figure*}

\begin{figure}
    \centering
    \includegraphics[width=0.48\textwidth]{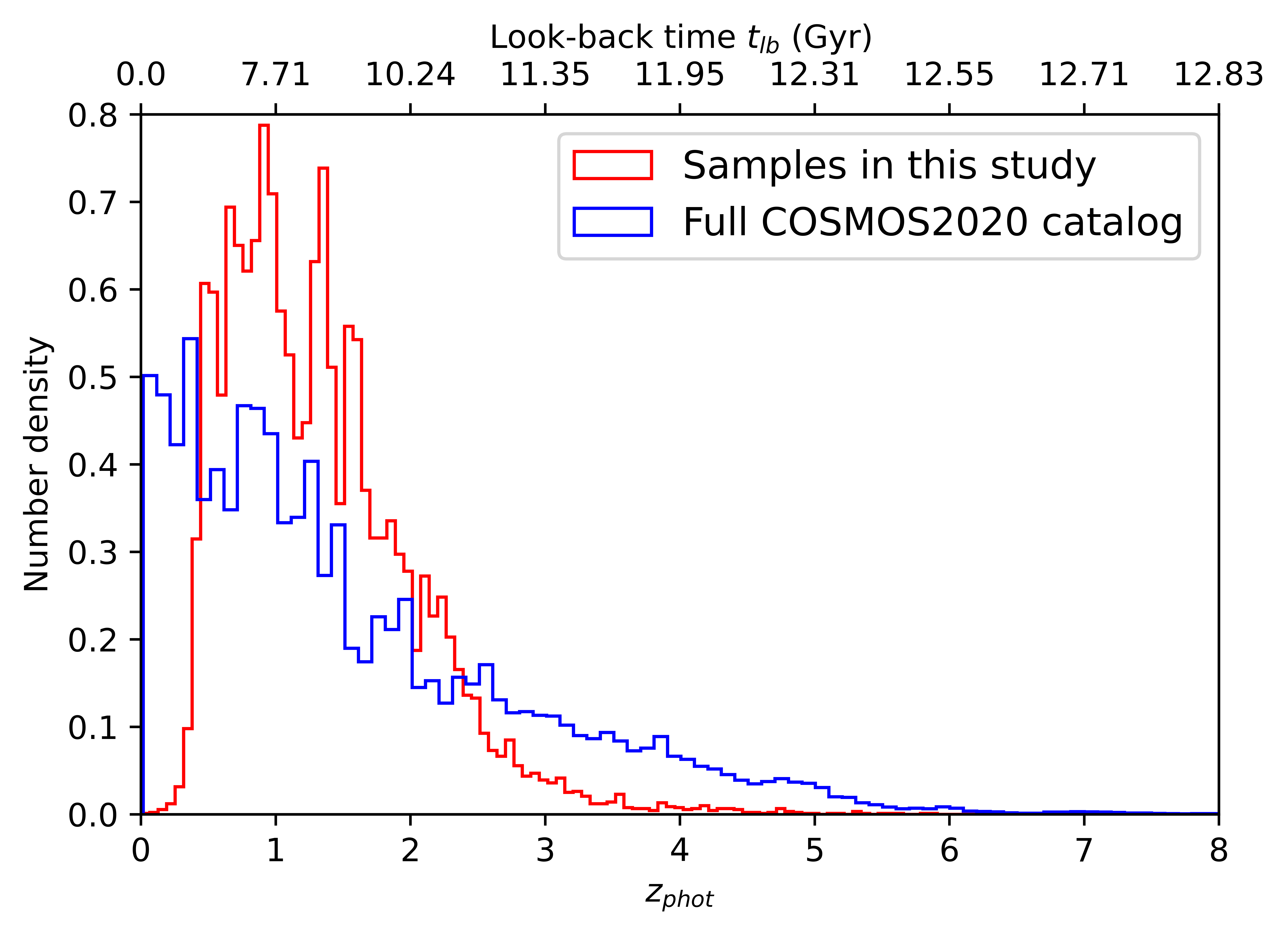}
    \caption{Normalized $z_{{\rm phot}}$ histogram for the galaxies. The red histogram indicates the distribution of photo-z's from \texttt{MAGPHYS+photo-z} for the 14,607 galaxies in our sample, while the blue histogram represents the parent distribution of the whole 964,506 galaxies from COSMOS2020 catalogue \citep{weaver22}, where the $z_{{\rm phot}}$ are derived using \texttt{LePhare} \citep{arnouts02, ilbert06}. We show the corresponding look-back time $t_{{\rm lb}}$ on the top axis.}
    \label{fig: z_phot histogram 14607}
\end{figure}

\texttt{MAGPHYS} fits the full SEDs of galaxies with known redshifts from the ultraviolet to the radio \citep{daCunha08, daCunha15} by combining the emission from stellar populations with the attenuation and re-emission of starlight by interstellar dust. The recent \texttt{MAGPHYS+photo-z} extension, described in \citet{battisti19}, extends the code to fit the SEDs of galaxies with unknown redshifts, and constrain the photometric redshift simultaneously with other galaxy physical properties. In practice, the code builds libraries of model UV-to-radio SEDs at different redshifts and compares them with the observed SEDs of galaxies, using a Bayesian method to obtain the likelihood distributions of physical parameters such as redshifts, stellar masses, and SFRs. There are two sets of libraries used in \texttt{MAGPHYS+photo-z}: (1) an optical library that describes emissions from stars, and (2) an infrared library that describes the emission from dust. The optical library uses the spectral population synthesis models of \citeg{bruzual&charlot03} and initial mass function from \citeg{chabrier03}; while the infrared library consists of models for PAHs and hot dust emitting in the mid-IR, and warm and cold dust components in thermal equilibrium that emit in the far-IR to submillimeter \citep{daCunha08}. These two sets of model libraries maintain the balance of the energy absorbed by dust (via attenuation in UV to near-IR) and the energy re-emitted by dust (via thermal emission in mid-IR to sub-mm). Due to insufficient models at $z<0.4$ to compare to based on the redshift prior that is adopted in the \texttt{MAGPHYS+photo-z}, galaxies with $z_{{\rm phot}} < 0.4$ are not constrained well. Therefore, we exclude galaxies with $z_{{\rm phot}} < 0.4$ after running the code \citep{battisti19}. An example \texttt{MAGPHYS+photo-z} fit is shown in Figure \ref{fig: MAGPHYS 1587}. We compare the distributions of photometric redshifts of our sample with those of the full COSMOS2020 sample in Figure \ref{fig: z_phot histogram 14607}. 
 
\begin{figure}
    \centering
    \includegraphics[width=0.48\textwidth]{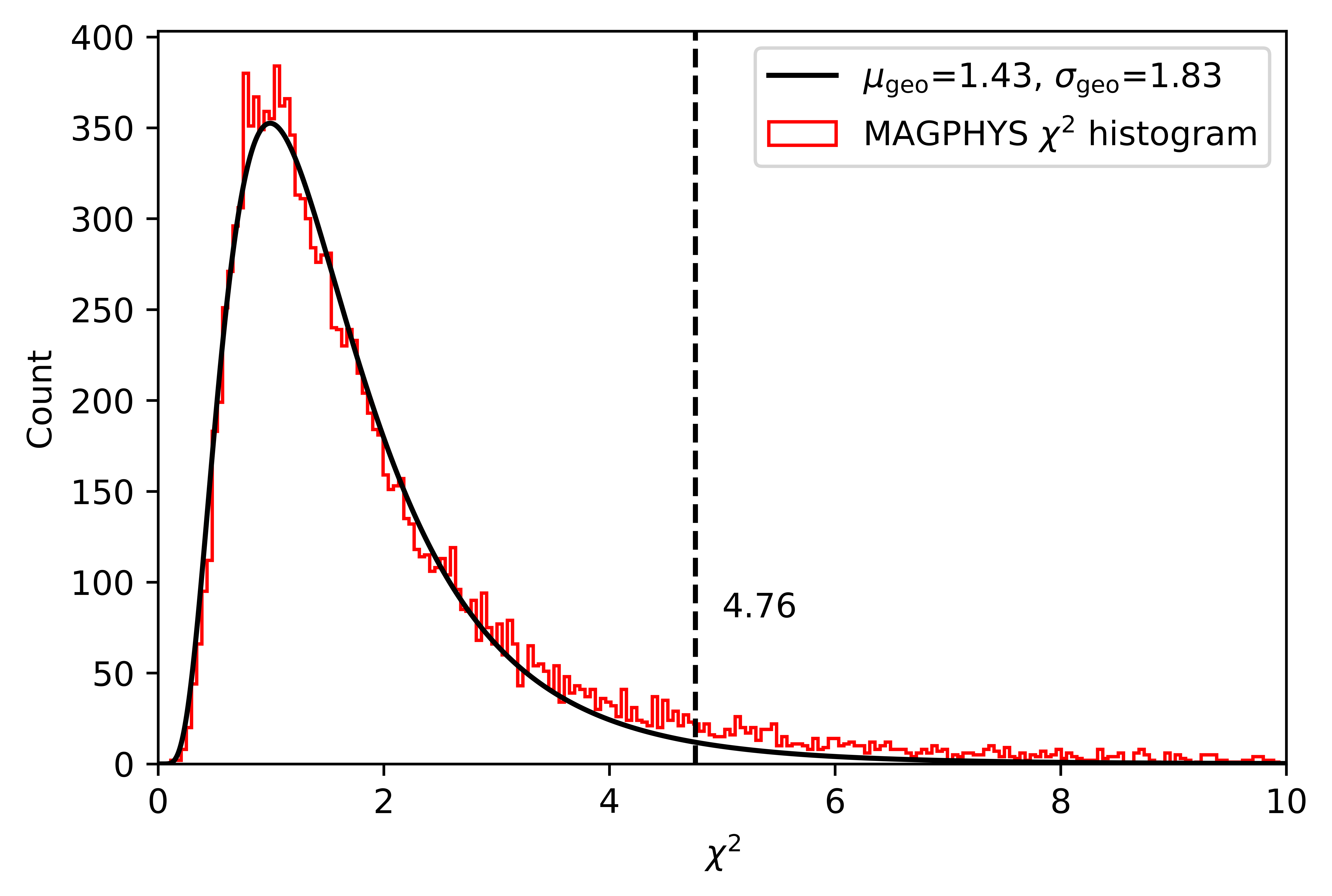}\\
    \caption{Distribution of \texttt{MAGPHYS+photo-z} fit $\chi^2$ of our 14,607 galaxies. The black curve represents the normalized lognormal distribution function fitting to the $\chi^2$ histogram, while the vertical black dashed line indicates the normal $2\sigma$ confidence cut within $\chi^2 \leq 4.76$.}
    \label{fig: chi^2 histogram 14607}
\end{figure}

We fit the SEDs of 14,607 galaxies with \texttt{MAGPHYS+photo-z} to determine the M*, SFR, $z_{{\rm phot}}$ and respective errors. We use the $\chi^2$ value of the best-fit model from \texttt{MAGPHYS+photo-z} as an indicator of the goodness of fit. We fit the $\chi^2$ distribution with a lognormal function (see Figure \ref{fig: chi^2 histogram 14607}) and convert the lognormal parameters $\mu$ and $\sigma$ to the geometric parameters $\mu_{\rm{geo}}$ and $\sigma_{\rm{geo}}$. Finally, we perform a $2\sigma$ confidence cut (i.e., $\chi^2<\mu_{\rm{geo}}+2\sigma_{\rm{geo}}$) to the histogram and remove the high-$\chi^2$ cases. The remaining galaxies are reduced to 13,639, and their $\chi^2 \leq 4.76$.  

However, some galaxies have problematic SEDs due to inconsistencies in fluxes and/or upper limits between bands. In these cases, \texttt{MAGPHYS+photo-z} derives large uncertainties of $z_{{\rm phot}}$ and distance-dependent parameters. In addition, some cases have multiple redshift solutions (e.g., degeneracies in Lyman vs Balmer break position), which can lead to multi-peaked solutions for distance-dependent derived properties. We adopt the following selection criteria based on key parameters (i.e., $z_{{\rm phot}}$, $M_*$ and SFR): 
$$
\begin{cases}
\sigma(z_{{\rm phot}}) &\leq 0.25 \\
\sigma(\log(M_*/M_{\odot})) &\leq 0.3 \\
\sigma(\log({\rm SFR}/M_{\odot}{\rm yr}^{-1})) &\leq 0.3, \\
\end{cases}
\label{err selection}
$$
where $\sigma(z_{{\rm phot}})$, $\sigma(\log(M_*/M_{\odot}))$ and $\sigma(\log({\rm SFR}/M_{\odot}{\rm yr}^{-1}))$ are measurement uncertainties for $z_{{\rm phot}}$, $M_*$ and SFR, respectively. The measurement uncertainty is calculated by half of difference between upper and lower 1$\sigma$ ($68\%$) boundary of probability distribution function (PDF) for each parameters derived by \texttt{MAGPHYS+photo-z}. We restrict the measurement uncertainty on redshift based on the size of our adopted redshift bins of 0.5 dex (i.e., 2 times larger than the uncertainty boundary). The limits of uncertainties on $M_*$ and SFR are set to 0.3 dex (roughly a factor of 2) because we want the measurement uncertainties to be lower than the typical intrinsic MS scatter, which is $\sim$0.3 dex \citeg{daddi07, ciesla14, speagle14}. These cuts remove $201$ galaxies ($1.5\%$) from our sample and we are left with $13,418$ galaxies.

\subsubsection{Reference Main Sequence Relation}
\label{Reference Main Sequence Relation} 

Numerous studies have examined the nature of the galaxy MS \citeg{speagle14, johnston15, tomczak16, pearson18, bisigello18, leslie20, thorne21}. \citet{tomczak16} and \citet{leslie20} introduce nonlinear fits to the MS. For our reference MS relation, we adopt \cite{leslie20} which also used galaxies in the COSMOS field, which has the form:  
\begin{center}
\begin{equation}
\begin{split}
\log({\rm SFR}(M, t)/M_{\odot}{\rm yr}^{-1}) &= S_0-a_1t-\log\left(1+\frac{10^{M_t'}}{10^M}\right) \\
M_t' &= M_0+a_2t,
\label{Leslie's equation}
\end{split}
\end{equation}
\end{center}

where $S_0 = 2.97^{+0.08}_{-0.09}$, $M_0 = 11.16^{+0.15}_{-0.16}$, $a_1 = 0.22^{+0.01}_{-0.01}$, $a_2 = 0.12^{+0.03}_{-0.02}$, $M$ is $\log(M^*/M_{\odot})$ and $t$ is the age of the universe in Gyr. \citet{leslie20} separate their sample into two classes, `All' and `SF' (`Star-forming'). We adopt the `SF' relation, which should coincide more closely with the sample used in our study. The \citet{leslie20} `SF' sample applies a colour selection (NUV-r-J cut) that will exclude `passive' galaxies with low SFRs, which has a similar role as our selection criterion described in next paragraph (see Section \ref{sSFR Selection}). The probed steller mass range in \citet{leslie20} is $9.0 \lesssim M_* \lesssim 11.0$ and redshift range is $0.3<z<6$. They use radio data to derive SFRs, which provides a dust-unbiased measurement of the SFR \citep{leslie20}. 

Although the goal of this study is not to investigate the relation between SFR and $M_*$, we note that the exact functional form of the MS is still under debate \citeg{katsianis21, leja22}. Different methods of estimating SFRs are thought to be the primary reason for differences between studies \citep{katsianis20}. Hence, despite using similar catalogues from the COSMOS field as \citet{leslie20} that use radio continuum for robust SFRs (dust-insensitive), there are some other systematic problems that can arise, such as priors, metallicities, timescales, stellar masses, ages, etc. We stress that the reference MS we show is intended only to guide the eye and we do not use it for any selection cuts (i.e., to define `on' vs `off' the MS), which instead are based on sSFR (see Section \ref{sSFR Selection}). Therefore, the choice of the reference MS has no impact on the results of this study. 

\subsubsection{\texorpdfstring{\MakeLowercase{s}{S}}SFR Selection}
\label{sSFR Selection} 

\begin{figure}
    \centering
    \includegraphics[width=0.48\textwidth]{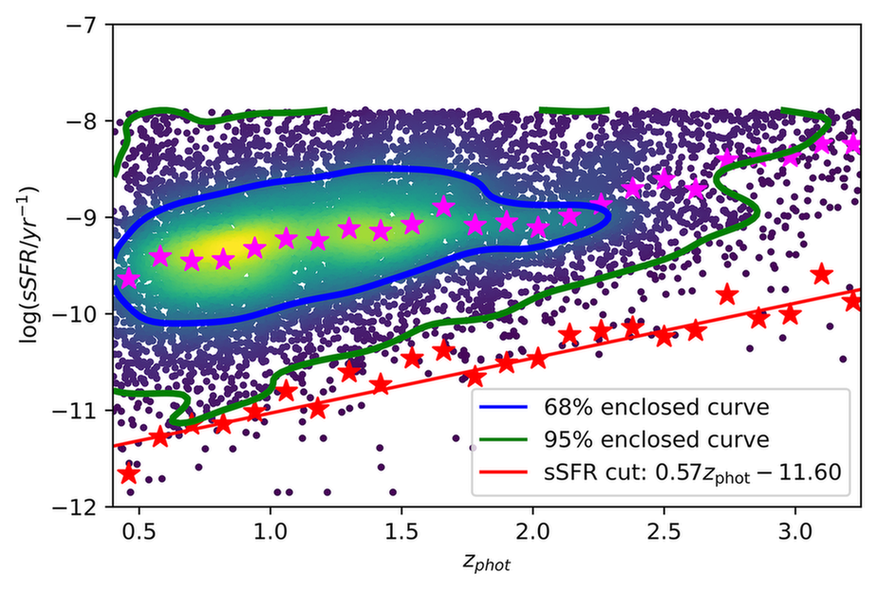}\\
    \caption{$\log({\rm sSFR})$ vs z enclosed contour plot for the 13,071 selection galaxies within $2\sigma-\chi^2$ from redshifts 0.4 to 3.25. The colours ranging from blue to yellow indicate the increasing number density. The number of galaxies inside the enclosed blue and green curve is $68\%$ ($1\sigma$) and $95\%$ ($2\sigma$) of the total population, respectively. The magenta points indicate the median value of $\log({\rm sSFR}/{\rm yr}^{-1})$, while the red points are the median-3$\sigma$ values for 24 sSFR bins. The red line is the sSFR cut adopted in this study, and there are 64 galaxies identified as quenched galaxies. Due to the minimum timescale of star formation in \texttt{MAGPHYS+photo-z}, there is a maximum value of $\log({\rm sSFR}/{\rm yr}^{-1})\sim-8$, corresponding to the adopted SFR timescale (100Myr), showing as a horizontal boundary in the diagram. }
    \label{fig: sSFR cut}
\end{figure}

We also adopt a specific-SFR (sSFR = SFR/$M_*$) cut to eliminate quenched galaxies (see the comparison to U-V-J selection in Appendix \ref{sSFR selection vs U-V-J Cut}). These `passive' galaxies form stars at a much lower rate for a given stellar mass compared to SF galaxies \citep{renzini15}. By definition, quenched galaxies have low sSFR values. The purpose of the sSFR cut is to remove these red galaxies to avoid overestimating the MS scatter. Since the sSFR of SF galaxies evolves with cosmic time \citep{madau&dickinson14}, we adopt a redshift-dependent cut\footnote{We explored adopting a constant cut at $\log({\rm sSFR}/{\rm yr}^{-1}) = -11$, which increases our sample by $\sim$100 galaxies, but this has a very small difference on our results. We suspect that this phenomenon could be driven by IR-selection, which is described in the next subsection. } in this selection criterion. In Figure \ref{fig: sSFR cut}, we use a linear regression model fit to the median-3$\sigma$ values for sSFR bins (24 bins) vs $z_{\rm phot}$ and remove quenched galaxies, which we define as $3-\sigma$ outliers lying below the equation: 

\begin{equation}
    \log({\rm sSFR}/{\rm yr}^{-1}) = 0.57z_{{\rm phot}}-11.60.
    \label{eq:sSFR_vs_z}
\end{equation}

\subsubsection{Influence of IR-selection on SFR- and mass-completeness}
\label{Influence of IR-selection on SFR- and mass-completeness}
\begin{figure}
    \centering
    \includegraphics[width=0.48\textwidth]{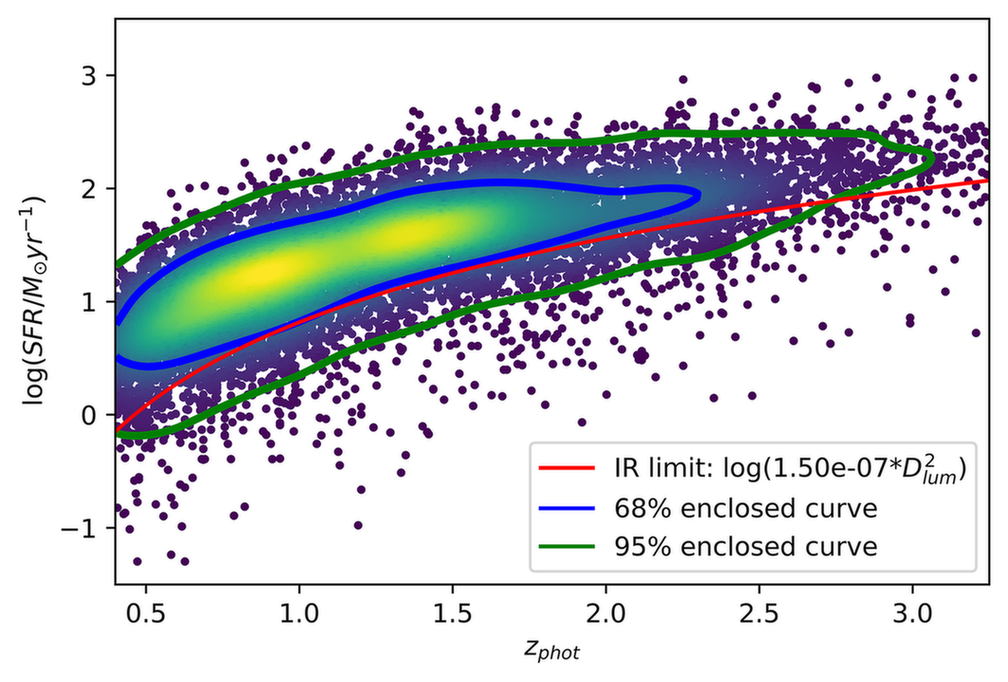}\\
    \caption{$\log({\rm SFR})$ vs z enclosed contour plot for the 13,071 selection galaxies within $\chi^2$ selection from redshifts 0.4 to 3.25. The equation \ref{eq: SFR IR-selection} is plotted as the red curve in this diagram.}
    \label{fig: IR-selection}
\end{figure}

A galaxy's SFR scales with the IR luminosity ($L_{{\rm IR}}$) \citep{kennicutt&evans12}. Due to this, the IR-selection criteria in our sample only includes galaxies above a certain SFR (depending on redshift), introducing an SFR bias. $L_{{\rm IR}}$ in this paper represents the integrated dust emission from both dust components in \texttt{MAGPHYS+photo-z} over all wavelengths. As the luminosity distance ($D_{{\rm lum}}$) increases, the lowest SFR of the SF galaxies we can observe will increase correspondingly\footnote{This excludes the negative-k correction effect}. Hence, the functional form for IR-selection in SFR is similar to the relationship between luminosity and redshift:  
\begin{equation}
    \begin{split}
    \log({\rm SFR}_{{\rm IR}}/M_{\odot}{\rm yr}^{-1}) &\propto \log(L_{{\rm IR}})\\ 
    &= \log(4\pi F_{{\rm IR}} D_{{\rm lum}}^2)\\ 
    &= \log(\alpha D_{{\rm lum}}^2),
    \label{eq:SFR_vs_Dlum}
    \end{split}
\end{equation}
where $\alpha$ is a constant factor determined by the data and $D_{{\rm lum}}$ is the luminosity distance in units of Mpc. By converting $D_{{\rm lum}}$ to $z_{{\rm phot}}$ and applying Equation \ref{eq:SFR_vs_Dlum} to $\log({\rm SFR})\ vs\ z_{{\rm phot}}$, we obtain an empirical estimate of our SFR limit with redshift based on the 1$\sigma$ lower boundary of our population, and the constant factor of the function $\alpha = 1.50\times10^{-7}$ corresponds to the lower boundary of the $68\%$ ($1\sigma$) population enclosed curve (see Figure \ref{fig: IR-selection}):  
\begin{equation} 
    \log({\rm SFR}_{{\rm IR}}/M_{\odot}{\rm yr}^{-1}) = \log(1.50\times10^{-7}(D_{{\rm lum}}/{\rm Mpc})^2).
    \label{eq: SFR IR-selection}
\end{equation}

 Due to the detection limits, we cannot trust our ability to detect galaxies that are below Equation \ref{eq: SFR IR-selection} in Figure \ref{fig: IR-selection}. This SFR incompleteness translates to an incompleteness on stellar mass (via galaxy MS relation). We infer the corresponding mass-completeness threshold at each redshift using the \citet{leslie20} MS relation. For subsequent analysis, we will refer to samples above and below this threshold as our mass-complete and mass-incomplete samples, respectively.

\section{Results and Analysis}
\label{Results and Analysis}

\subsection{The Role of Redshift uncertainty and IR data on the Measured Scatter of the MS}
\label{The Role of Redshift uncertainty and IR data on the Measured Scatter of the MS}

In this study, we use \texttt{MAGPHYS+photo-z} to constrain the stellar masses and SFRs of our galaxies because it uses the full wavelength range from UV to radio, and it constrains the photometric redshifts jointly with the other physical parameters. Using the full SED provides the tightest possible constraints on $M_*$ and SFR, thus minimizing the main sequence scatter that is due to errors on these parameters. Obtaining the photometric redshift at the same time allows us to fold in the redshift error into the errors on $M_*$ and SFR. This improves our ability to quantify the `observational' scatter on the main sequence and, in turn, characterise its intrinsic MS scatter. In this section, we test the accuracy of \texttt{MAGPHYS+photo-z} and quantify the influence that the redshift precision and inclusion of IR data have on derived physical properties (for the COSMOS filter set). 

\subsubsection{Accuracy of $z_{{\rm phot}}$ relative to $z_{{\rm spec}}$ } 
\label{Accuracy of z_phot relative to z_spec}

To examine the $z_{{\rm phot}}$ accuracy of \texttt{MAGPHYS+photo-z}, we use the latest COSMOS master spectroscopic catalogue (curated by M. Salvato for internal use within the COSMOS collaboration), which is the same dataset used to originally test the code \citep{battisti19}. There are spectroscopic redshifts, $z_{{\rm spec}}$, for 3,873 out of the 14,607 galaxies in our sample. After applying the $\chi^2$ cut, we obtain 3,724 galaxies. Here we adopt some metrics defined in Section 4.1.1 of \citet{battisti19} to estimate the accuracy of $z_{{\rm phot}}$. We find $\sigma_{\rm{NMAD}} = 0.086$, $\eta = 4.2\%$ and $z_{\rm{bias}} = -0.002$, where $\sigma_{\rm{NMAD}}$ (normalized median absolute deviation) is known as the precision or scatter of the data, $\eta$ characterises the fraction of catastrophic failures and $z_{\rm{bias}}$ represents the accuracy of the redshift (i.e. systematic deviation or bias). The value of $z_{\rm{bias}}$ is much smaller than $\sigma_{\rm{NMAD}}$, and hence we constrain the redshifts very well with the multiple UV to radio bands. Since we use a similar database as the one used in \citet{battisti19}, the results of $\sigma_{\rm{NMAD}}$, $\eta$ and $z_{\rm{bias}}$ should be similar. As a comparison, these values calculated in \citet{battisti19} are $\sigma_{\rm{NMAD}} = 0.032$, $\eta = 0.037$, $z_{\rm{bias}} = -0.004$ for the COSMOS2015 samples, respectively.  

\begin{figure*}
     \centering
    $$
    \begin{array}{cc}
    \centering
    \includegraphics[width=0.48\textwidth]{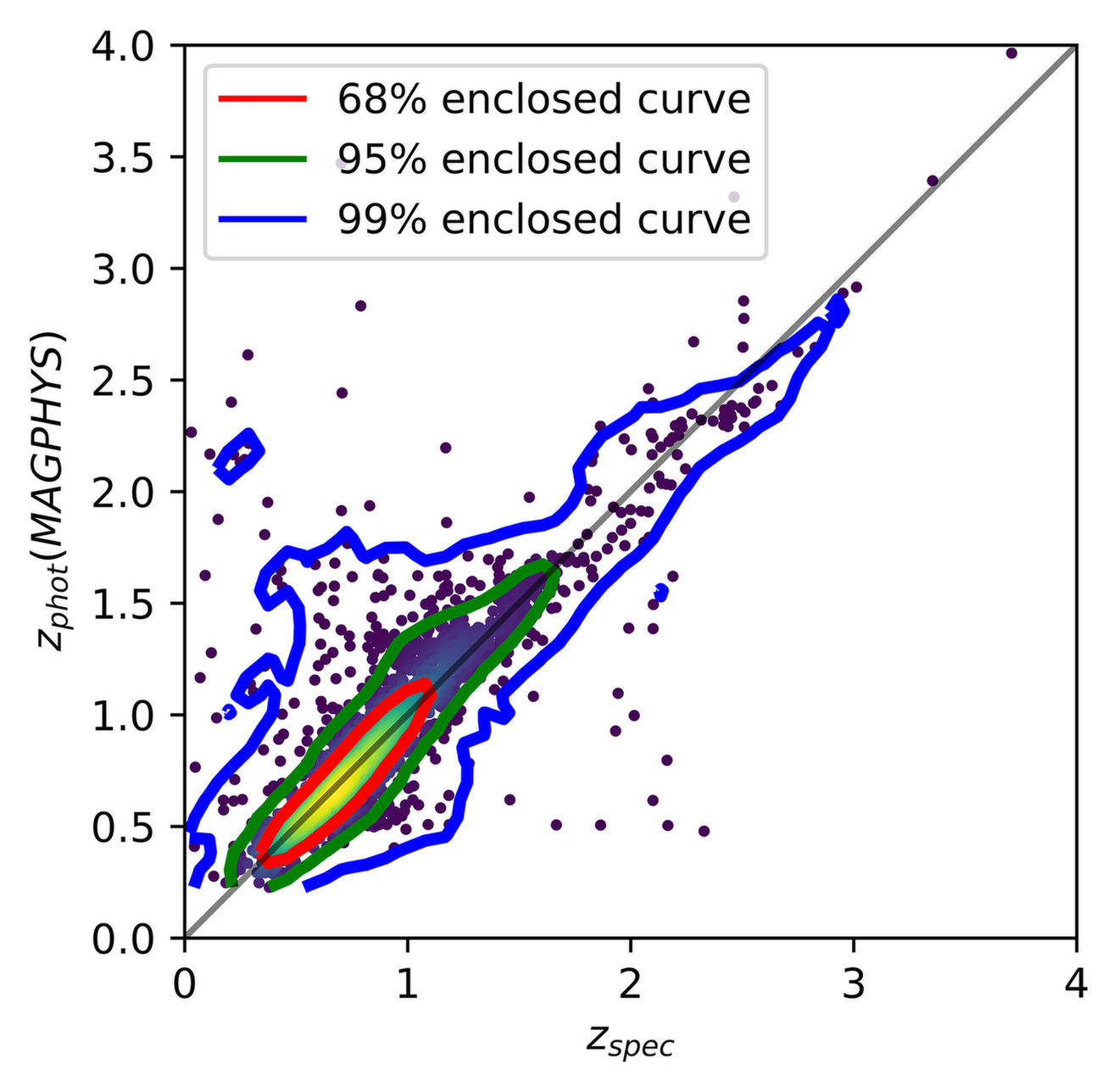} &
    \includegraphics[width=0.48\textwidth]{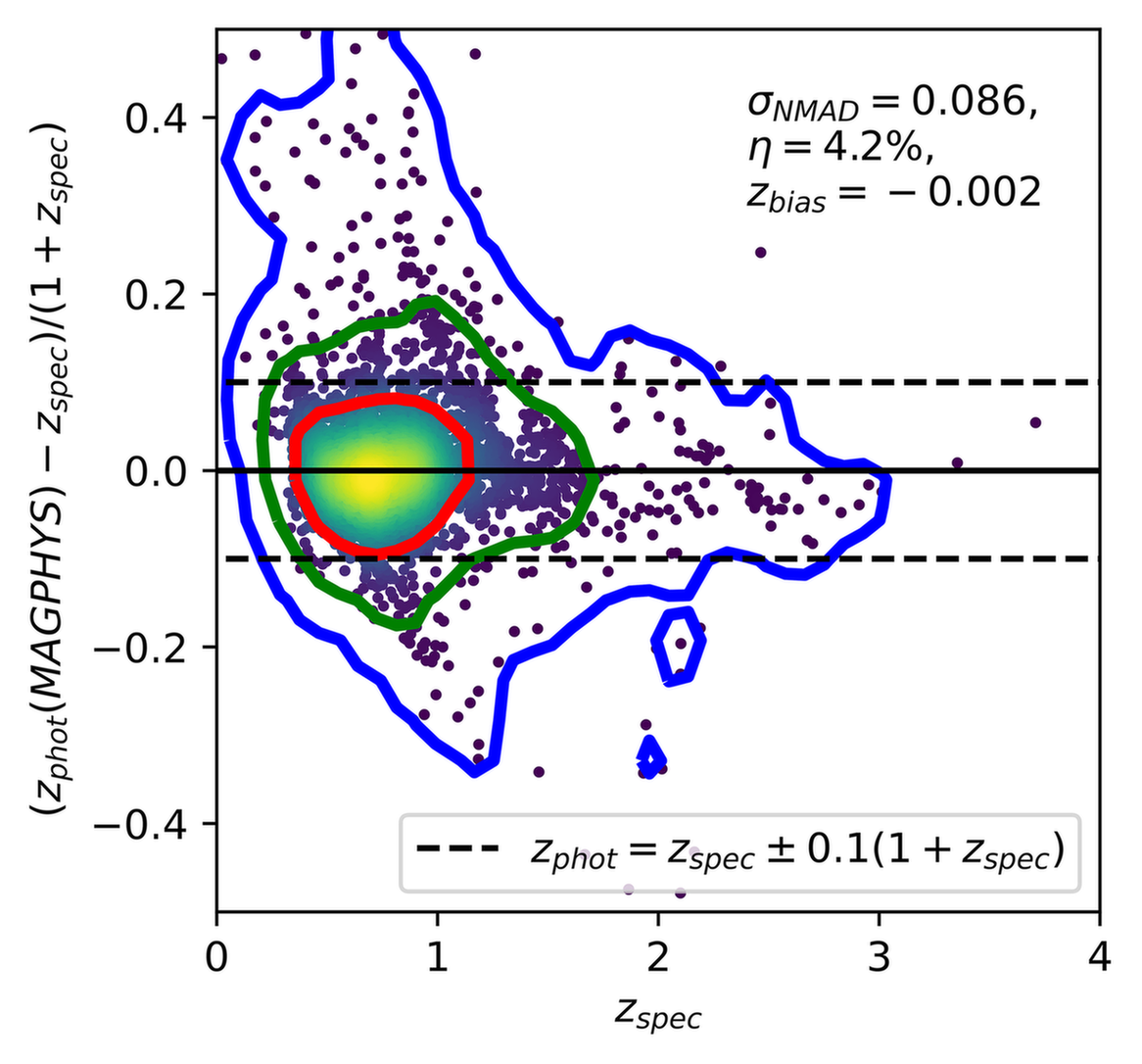} \\
    \includegraphics[width=0.48\textwidth]{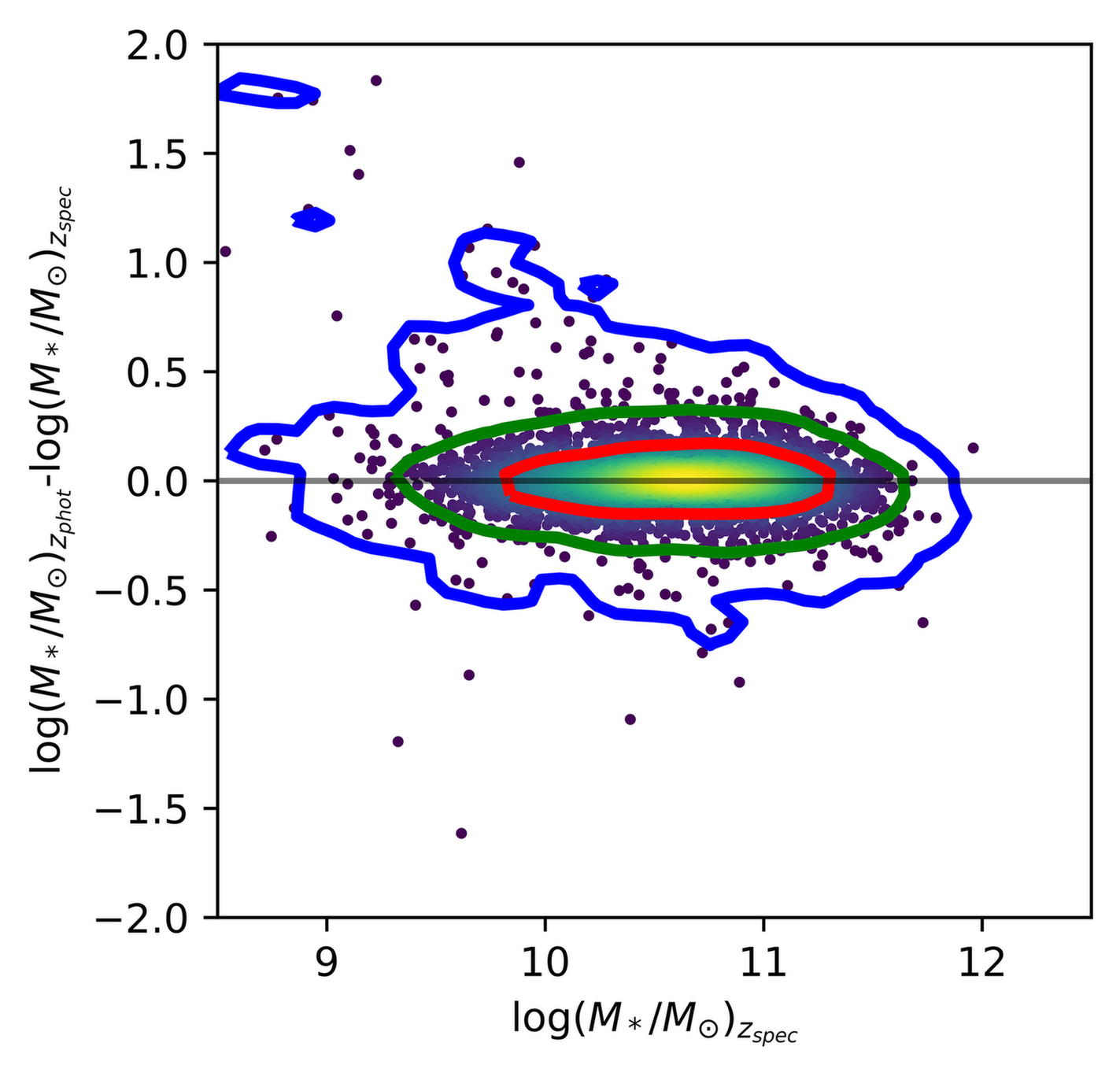} &
    \includegraphics[width=0.48\textwidth]{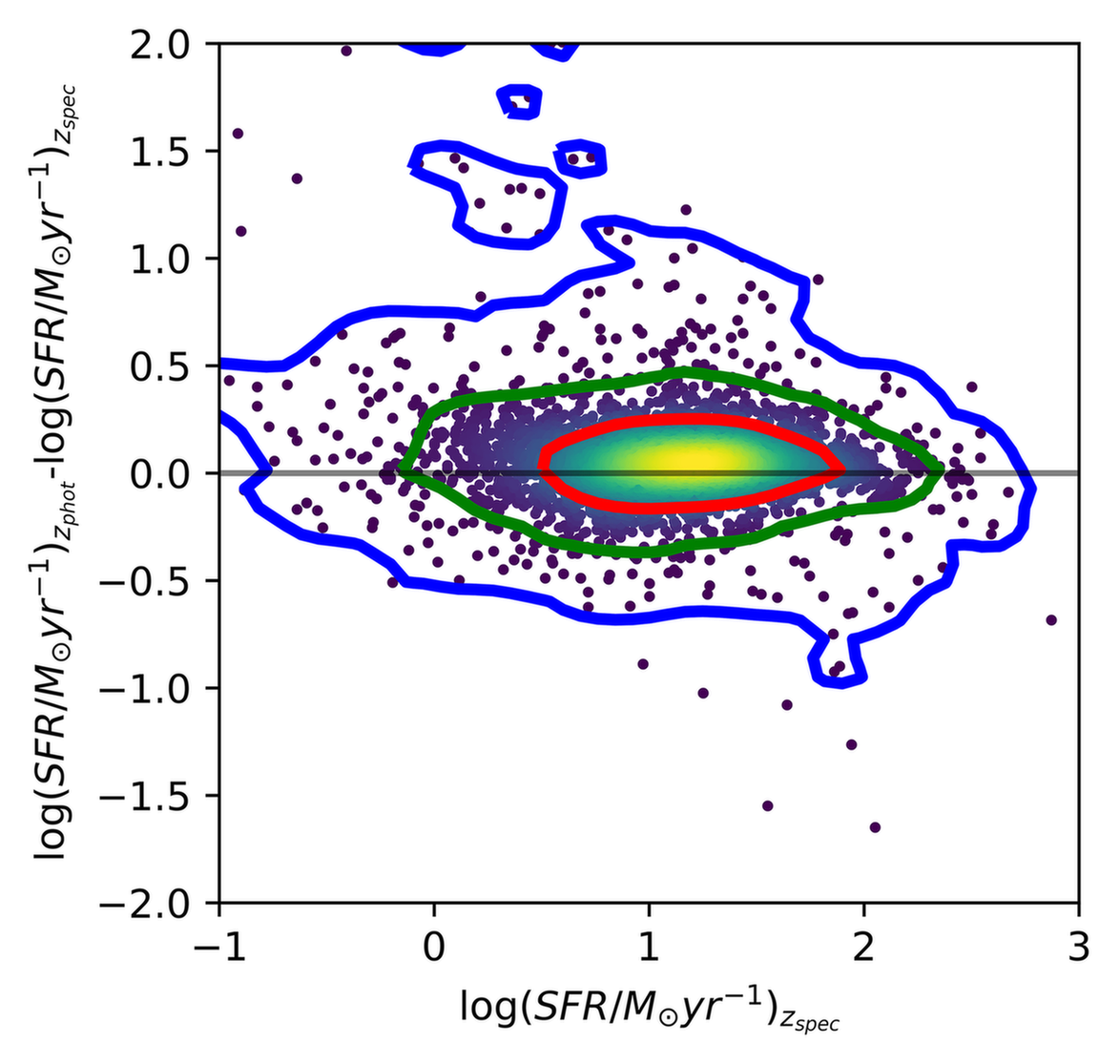} \\
    \end{array}
    $$
    \caption{Upper left panel: Comparison of measurement uncertainties between default MAGPHYS high-$z$ (i.e, fixed to $z_{{\rm spec}}$) and MAGPHYS+photoz runs for the subsample of 3,724 $\chi^2$-selection galaxies with spectroscopic redshifts. Upper right panel: redshift accuracy ($(z_{{\rm phot}}-z_{{\rm spec}})/(1 + z_{{\rm spec}})$) as a function of $z_{{\rm spec}}$. The redshift scatter ($\sigma_{\rm{NMAD}}$), catastrophic failure rate ($\eta$), and redshift bias (median($(z_{{\rm phot}}-z_{{\rm spec}})/(1 + z_{{\rm spec}})$)) values are shown at the upper right corner. Lower panels: Difference in $M_*$ and SFR derived by $z_{{\rm phot}}$ and $z_{{\rm spec}}$ as a function of the $z_{{\rm spec}}$-derived values. The 2D histogram/scatterplot colors range from blue to yellow with increasing number density. The black line in each sub-diagram is the one-to-one relation as reference; red, green and blue curves enclose $68\%$, $95\%$ and $99\%$ populations of sample galaxies within $2\sigma$-$\chi^2$ cut. }
    \label{fig: 4 panels}
\end{figure*}

The upper panels of Figure \ref{fig: 4 panels} is a demonstration of $z_{{\rm phot}}$ accuracy of \texttt{MAGPHYS+photo-z}, which also shows a comparison between the $M_*$ and SFR derived from $z_{{\rm phot}}$ and $z_{{\rm spec}}$. The median values of differences for $M_*$ and SFR are 0.00 and 0.05 dex, respectively, reflecting that \texttt{MAGPHYS+photo-z} does not affect the overall measurement of $M_*$ and SFR. Therefore, we do not expect that relying on $z_{{\rm phot}}$ will introduce significant bias or dominate the uncertainty of the other derived properties.  

\subsubsection{Uncertainties of $z_{{\rm phot}}$, $M_*$ and SFR}
\label{Uncertainties of z_phot, M_* and SFR}

\begin{figure*}
     \centering
    $$
    \begin{array}{ccc}
    \centering
    \includegraphics[width=0.31\textwidth]{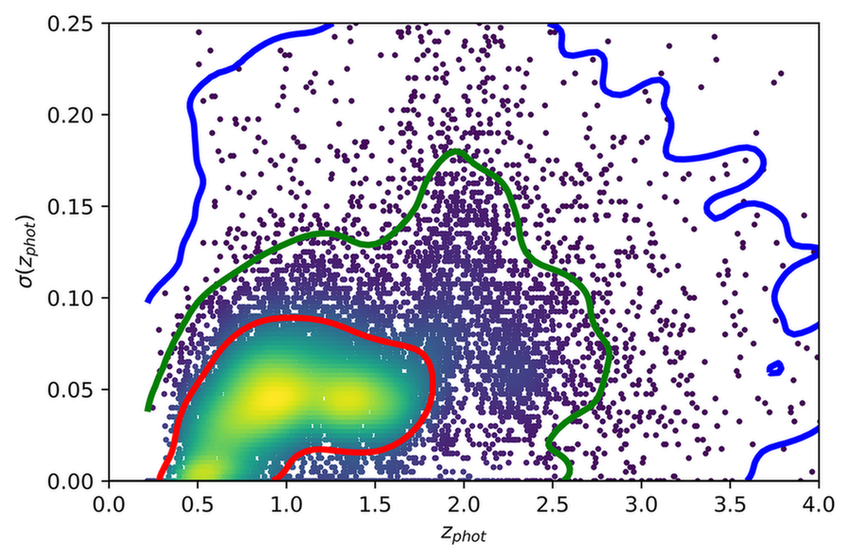} & 
    \includegraphics[width=0.31\textwidth]{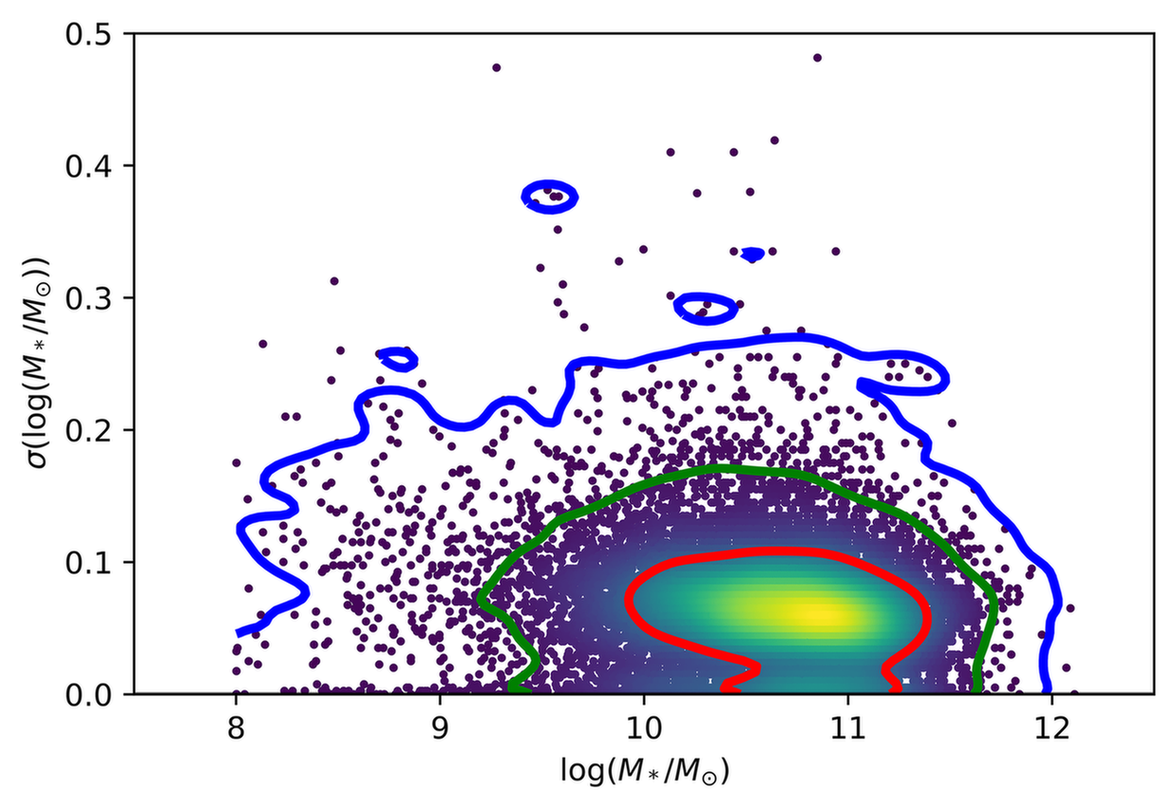} & 
    \includegraphics[width=0.31\textwidth]{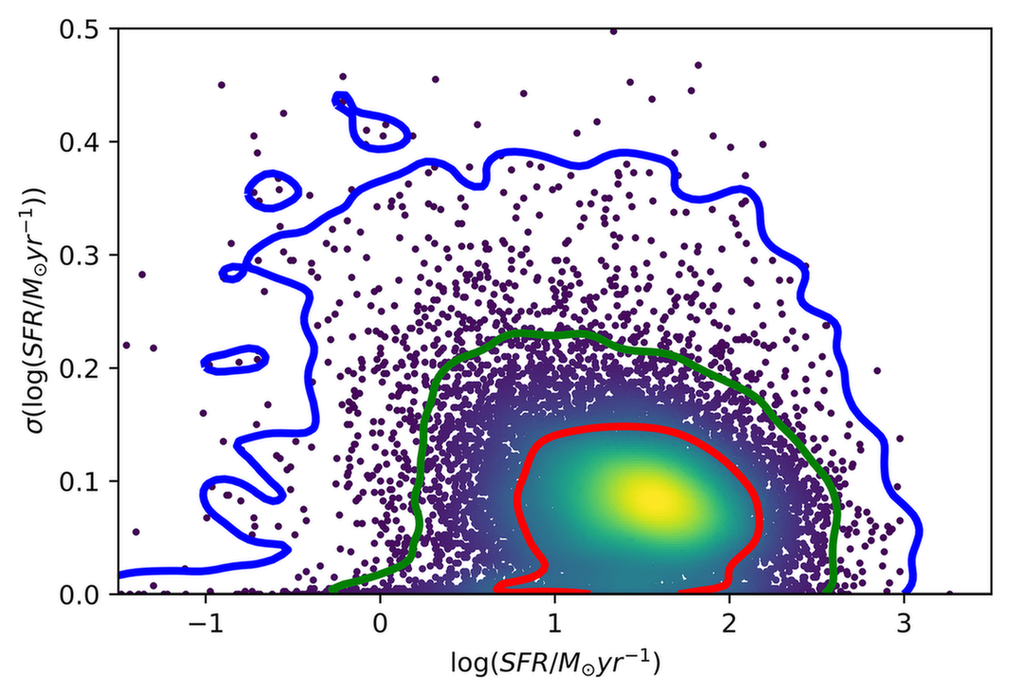} \\  
    \end{array}
    $$
    \caption{Distribution of the values and uncertainties for the key parameters of our study for the 13,418 galaxies in our sample. Red, green and blue curves enclose $68\%$, $95\%$ and $99\%$ populations of sample galaxies within $2\sigma$-$\chi^2$ cut. The median values of $z_{{\rm phot}}$, $\log(M_*/M_{\odot})$ and $\log({\rm SFR}/M_{\odot}{\rm yr}^{-1})$ are 1.20, 10.64 and 1.42, while the median values of uncertainties are 0.05, 0.06 and 0.08 dex, respectively. }
    \label{fig: S3.1.2}
\end{figure*}

We characterise the measurement uncertainties of $z_{{\rm phot}}$, $M_*$ and SFR for our sample of 13,418 galaxies in Figure \ref{fig: S3.1.2}. In our analysis, the data are separated into five bins of $M_*$ with a width of $\geq 0.5$ dex at a specific redshift epoch. The uncertainty in $M_*$ in each bin is $\sim 0.06$ dex, while the $z_{{\rm phot}}$ uncertainty is $\sim 0.05$. Both uncertainties are $\sim 10$ times smaller than the bin size in this study ($\geq 0.5$ dex for $\log(M_*)$ bin and $\sim 0.5$ for redshift epoch), therefore we do not anticipate that these uncertainties will have a substantial impact on the derived intrinsic scatter on the MS. The median value of SFR's uncertainty is 0.08 dex, which is comparable to the scatter of galaxies on the MS \citep[e.g., $\sim 0.2$ dex;][]{speagle14}. Thus, when measuring the intrinsic scatter of MS, we need to consider SFR's uncertainty as the component of the scatter in MS and remove it properly to obtain the intrinsic MS scatter. Our method for removing this component is described in Section \ref{Measuring the intrinsic MS scatter}.

\subsubsection{Contribution of including IR data to the Uncertainties of $M_*$ and SFR}
\label{Contribution of including IR data to the Uncertainties of M_* and SFR}

\begin{figure*}
     \centering
    $$
    \begin{array}{ccc}
    \centering
    \includegraphics[width=0.31\textwidth]{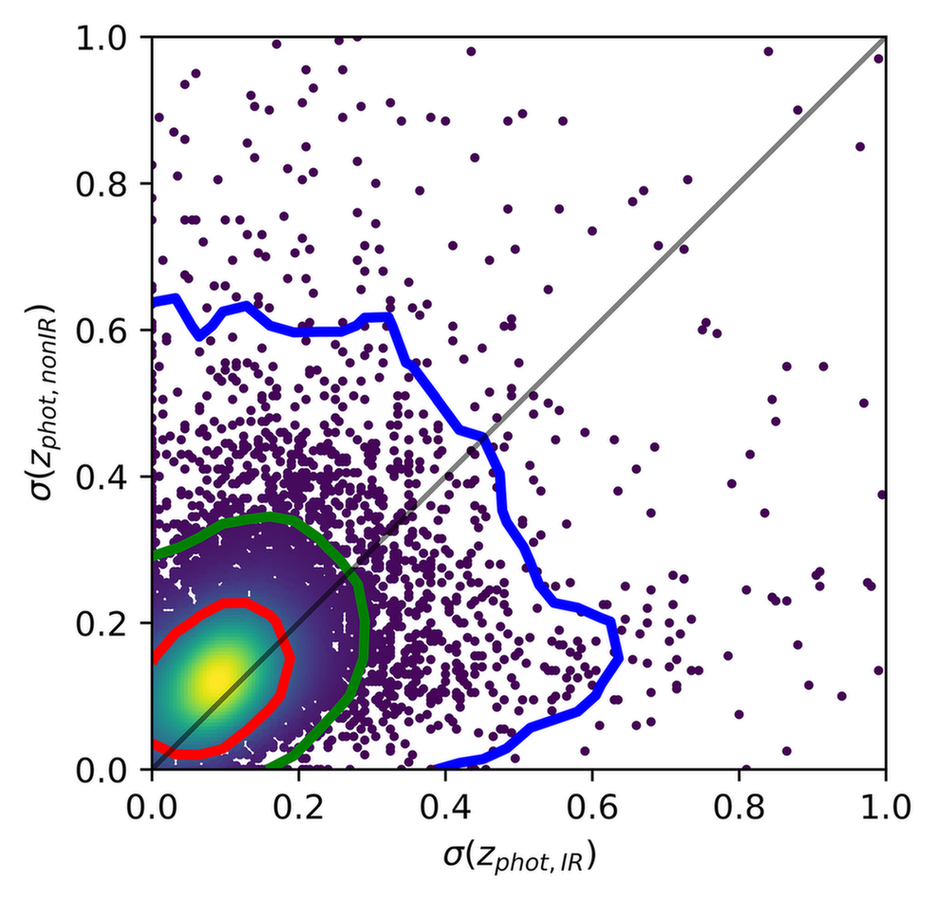} &
    \includegraphics[width=0.31\textwidth]{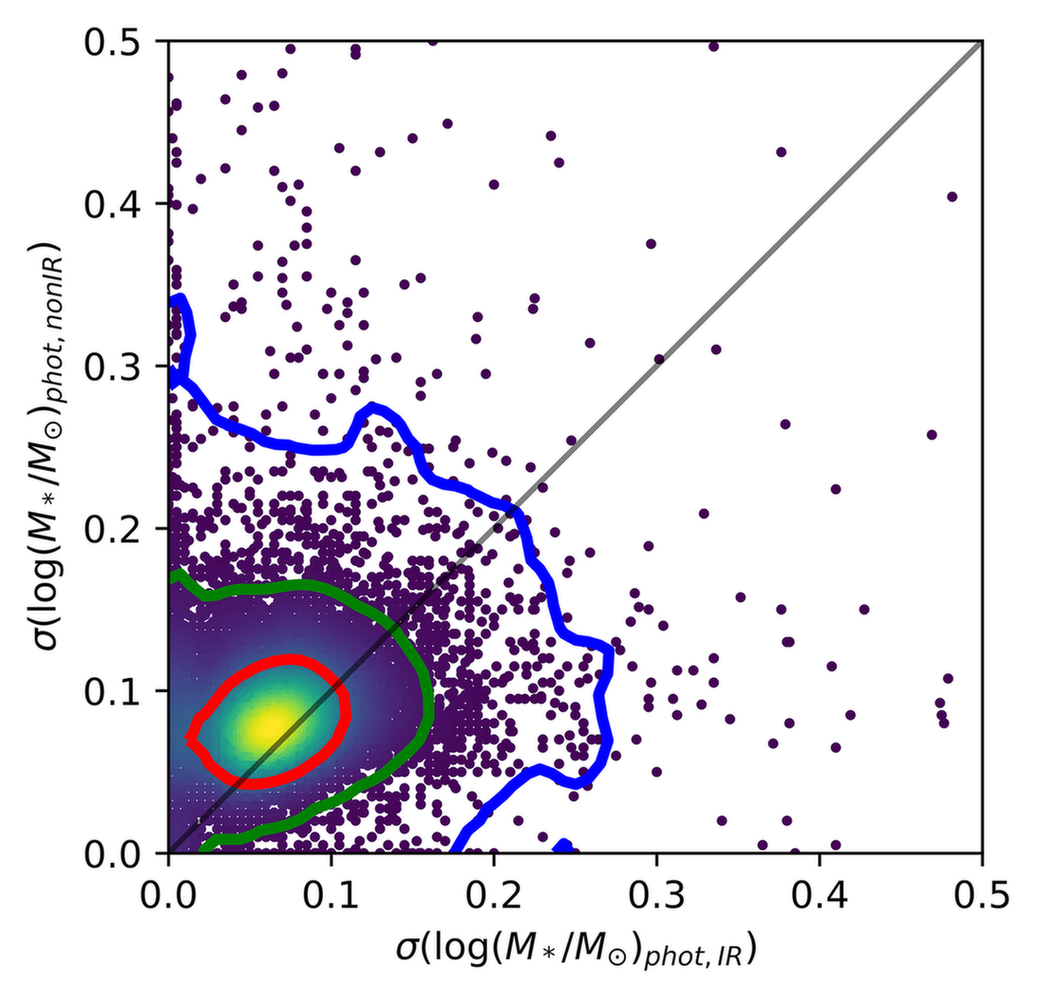} & 
    \includegraphics[width=0.31\textwidth]{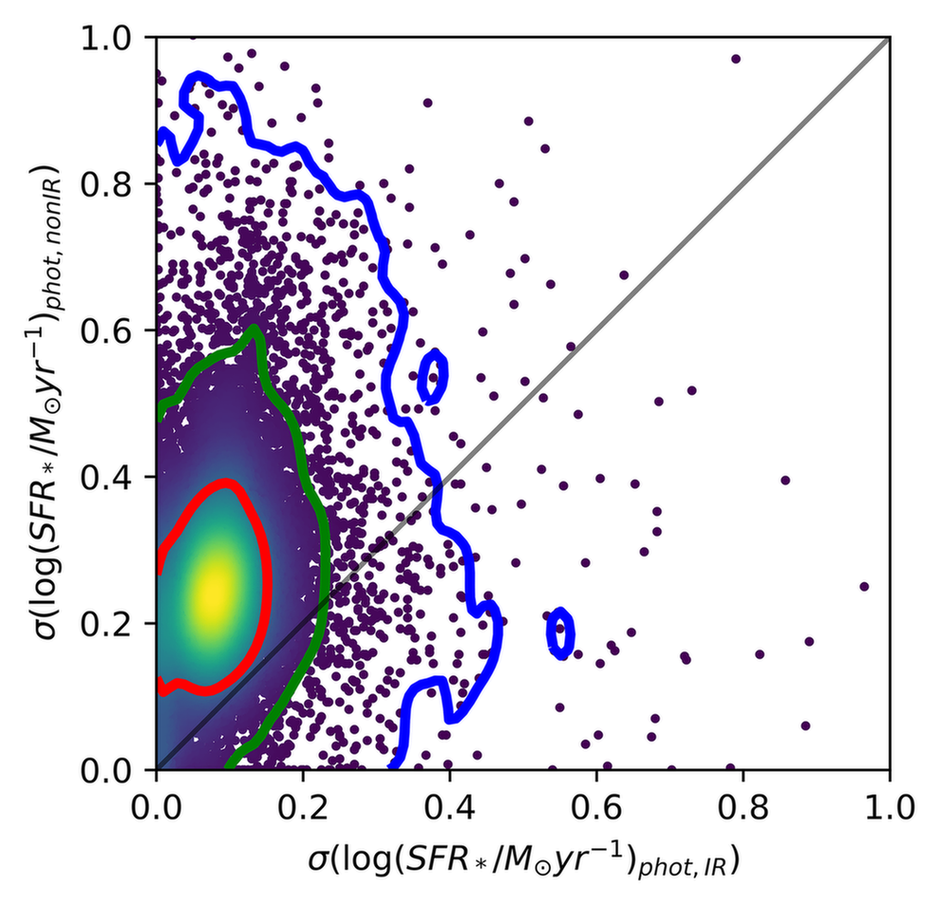} \\ 
    \includegraphics[width=0.31\textwidth]{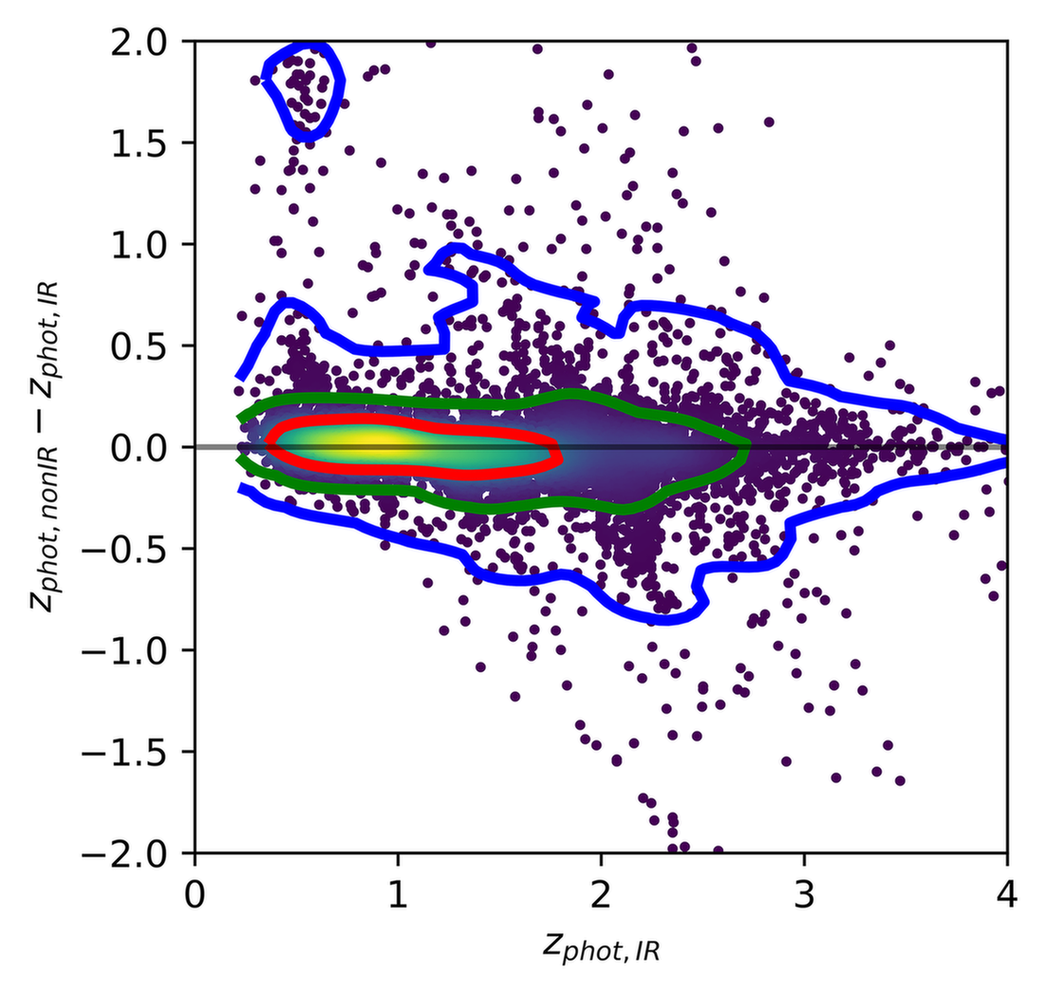} &
    \includegraphics[width=0.31\textwidth]{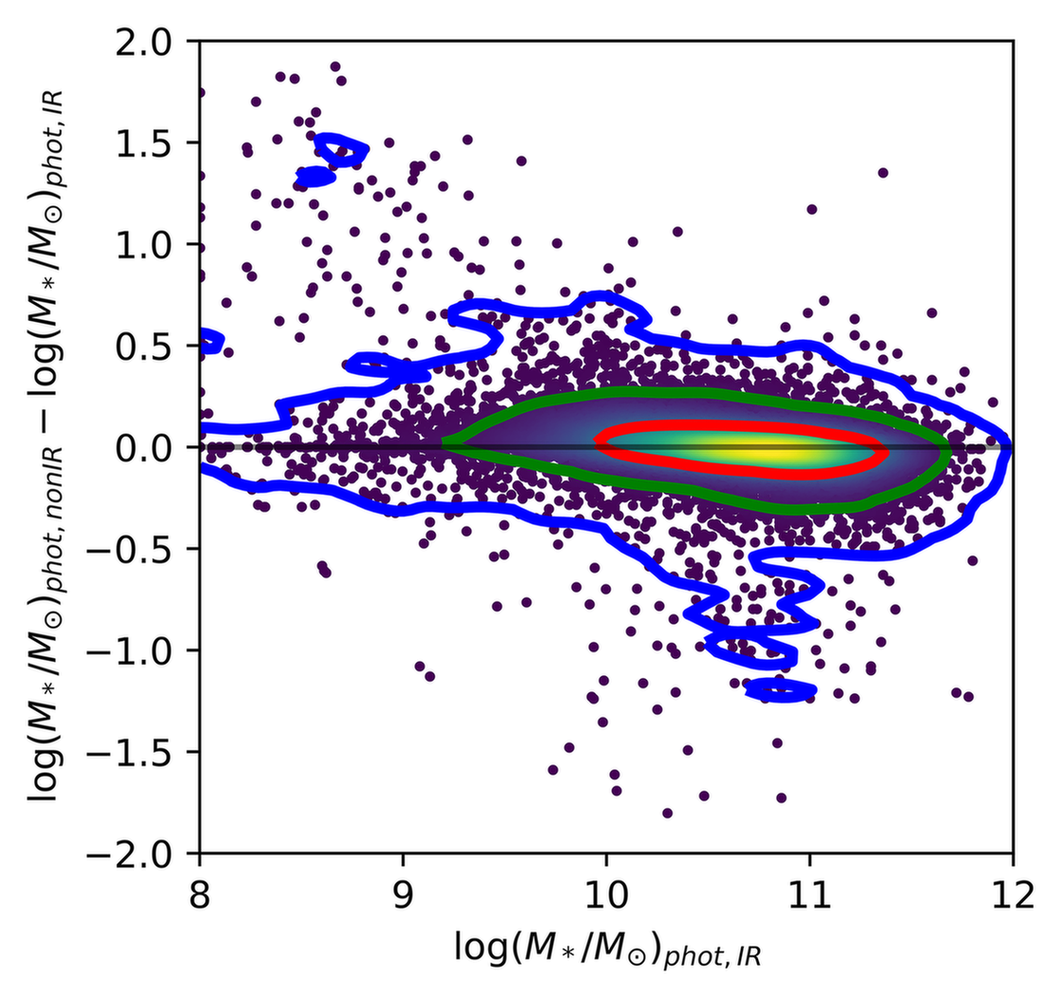} & 
    \includegraphics[width=0.31\textwidth]{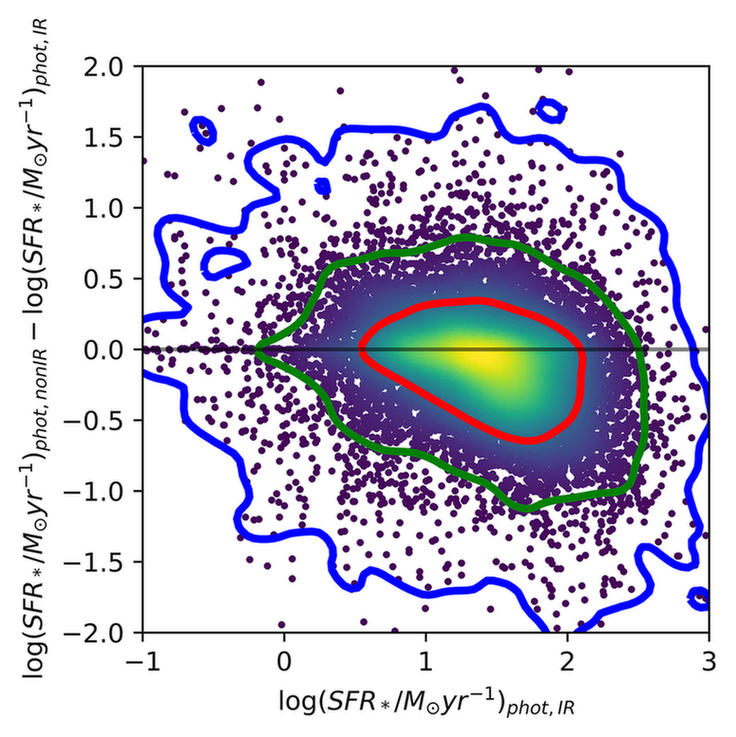} \\ 
    \end{array}
    $$
    \caption{Upper panels: Comparison between the uncertainties on $z_{{\rm phot}}$, $M_*$ and SFR for \texttt{MAGPHYS+photo-z} runs including IR bands (labeled `phot,IR') and without IR or radio bands (`phot,nonIR'). Differing from the cases for $z_{{\rm phot}}$ and $M_*$, the SFR's uncertainty increases dramatically for SED fits without the IR data. The median uncertainty values derived from \texttt{MAGPHYS+photo-z} including the IR bands (distributed along x-axis) are $0.05$, $0.06$ and $0.08$ dex for $z_{{\rm phot}}$, $M_*$ and SFR, while the median uncertainty values derived from non-IR runs (distributed along y-axis) are $0.08$, $0.08$ and $0.25$ dex, respectively. The black line in each sub-diagram is the one-to-one relation as reference; red, green and blue curves enclose $68\%$, $95\%$ and $99\%$ populations of sample galaxies within $2\sigma$-$\chi^2$ cut. Lower panels: Difference in measurements between the SED fits with and without IR bands included as a function of the measurements derived from fitting including IR bands. For the COSMOS2020 data, we find that the dominant factor in accurately constraining the scatter in the main sequence is whether the IR bands are included to constrain the dust-obscured SFR.}
    \label{fig: non-IR}
\end{figure*}

IR wavelengths probe dust emission and provide information regarding the amount of dust-obscured star formation. By excluding IR observations from the SED fits, we can determine the impact of these bands on the uncertainties of $M_*$ and SFR. We rerun the \texttt{MAGPHYS+photo-z} without fitting the observational data for filters at wavelengths longer than IRAC2 (4.5um) for the same 14,607 galaxies. After rejecting the cases with bad fits ($\chi_2 > 2\sigma$), we compare the uncertainties of $z_{{\rm phot}}$, $M_*$ and SFR derived from UV to near-IR photo-z fitting to those results from fitting the full available SED in the top panels of Figure \ref{fig: non-IR}. As expected \citep{battisti22}, the non-IR fits tend to come with larger measurement uncertainties because fewer observations are available to constrain the models. For $z_{\rm phot}$ and $M_*$, including the IR bands only leads to a relatively small improvement (i.e., decrease) in the uncertainty. In contrast, for SFR, the median uncertainty when IR bands are not included is nearly 2.5 times larger than that with the IR bands. This is because the IR bands are important to distinguish the amount of dust-obscured star formation. It is harder to accurately measure the intrinsic scatter of the MS with the larger measurement error in SFR. Therefore, by restricting the sample to sources where SED fitting can be performed that include IR filters, we significantly reduce the amount of scatter of the MS arising from measurement uncertainty to accurately constrain the intrinsic MS scatter. The lower panels of Figure \ref{fig: non-IR} show the difference in the values of $z_{{\rm phot}}$, $M_*$ and SFR with and without the IR bands included. It can be seen that the median of the difference remains close to zero as a function of each property suggesting that there is minimal bias occurring as a result of the \texttt{MAGPHYS+photo-z} priors. 

\subsection{Measuring the intrinsic MS scatter}
\label{Measuring the intrinsic MS scatter}

We divide our sample into 6 redshift bins: $z_{{\rm phot}} = 0.5,\ 1.0,\ 1.5,\ 2.0,\ 2.5\ \&\ 3.0$, with widths of $\Delta z=0.5$ except for the lowest bin, which spans 0.4-0.75 due to the limitation on the redshift prior for \texttt{MAGPHYS+photo-z} (Section \ref{MAGPHYS+photo-z}). At each redshift, we further divide the sample into 5 stellar mass bins. We determine the $\log({\rm SFR})$ dispersion (standard deviation) of the galaxies within each bin relative to the median $\log({\rm SFR})$ measurement uncertainties. We set the following five bins for all selected galaxies according to their mass: $\log(M_*)<9.5\log(M_{\odot})$, $9.5$ to $10.0\log(M_{\odot})$, $10.0$ to $10.5\log(M_{\odot})$, $10.5$ to $11.0\log(M_{\odot})$ and $>11.0\log(M_{\odot})$. The values adopted for each bin is the median $\log({\rm SFR})$ and $M_*$ of each group. We characterise the intrinsic MS scatter in the range of $0.4 < z_{{\rm phot}} < 3.25$ (see Figure \ref{fig: z_phot histogram 14607}). The upper boundary ($z_{{\rm phot}} = 3.25$) corresponds to where our sample size dramatically decreases such that we do not have enough sources to properly characterize the MS scatter. 
Then we take the further selections of $2\sigma$-$\chi^2$ and sSFR cut (see Section \ref{MAGPHYS+photo-z} and \ref{sSFR Selection}) to reduce the effects of quiescent galaxies on the intrinsic MS scatter.

\begin{figure*}
    \centering
    $$
    \begin{array}{ccc}
    \centering
    \includegraphics[width=0.48\textwidth]{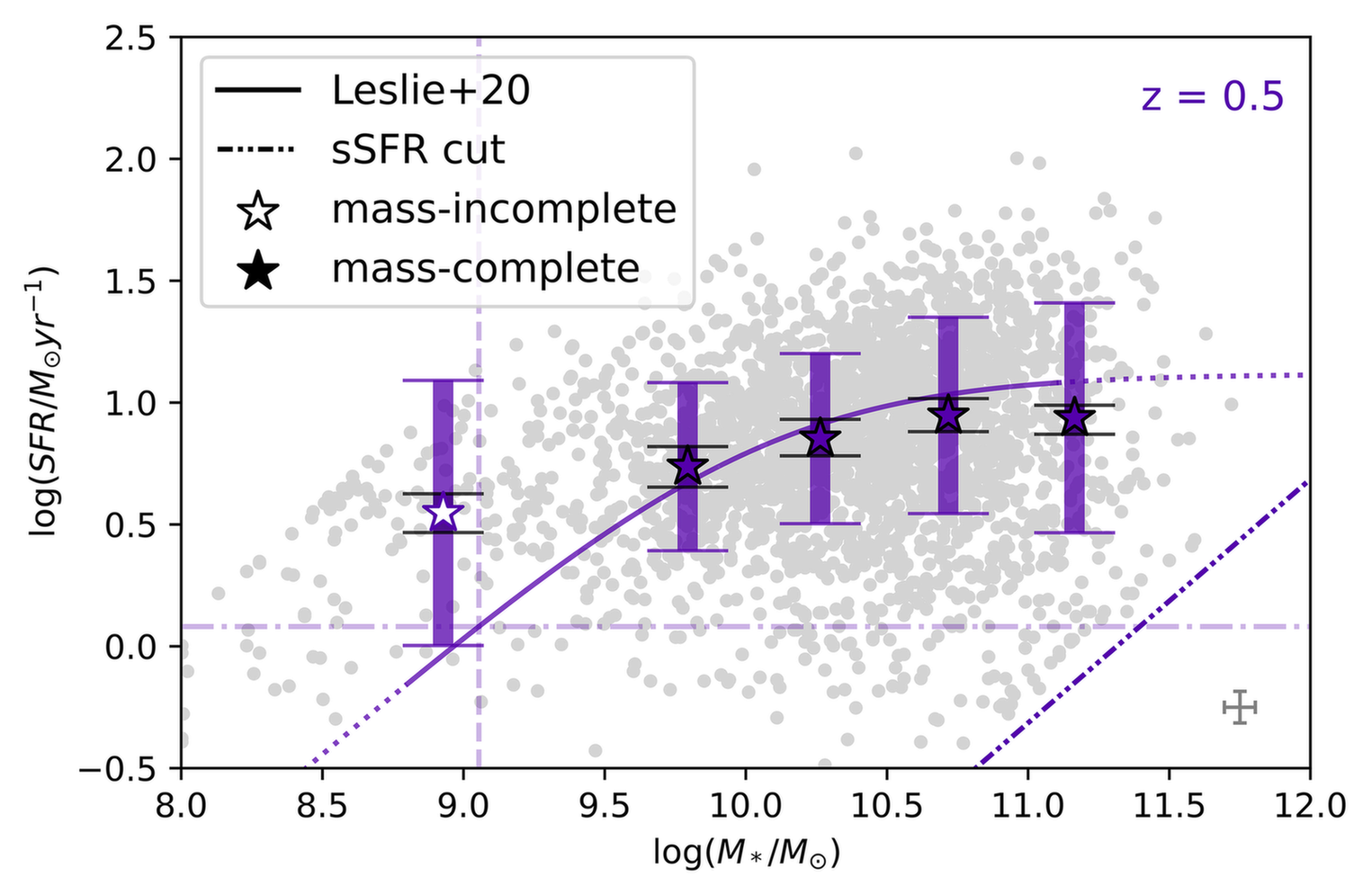} &
    \includegraphics[width=0.48\textwidth]{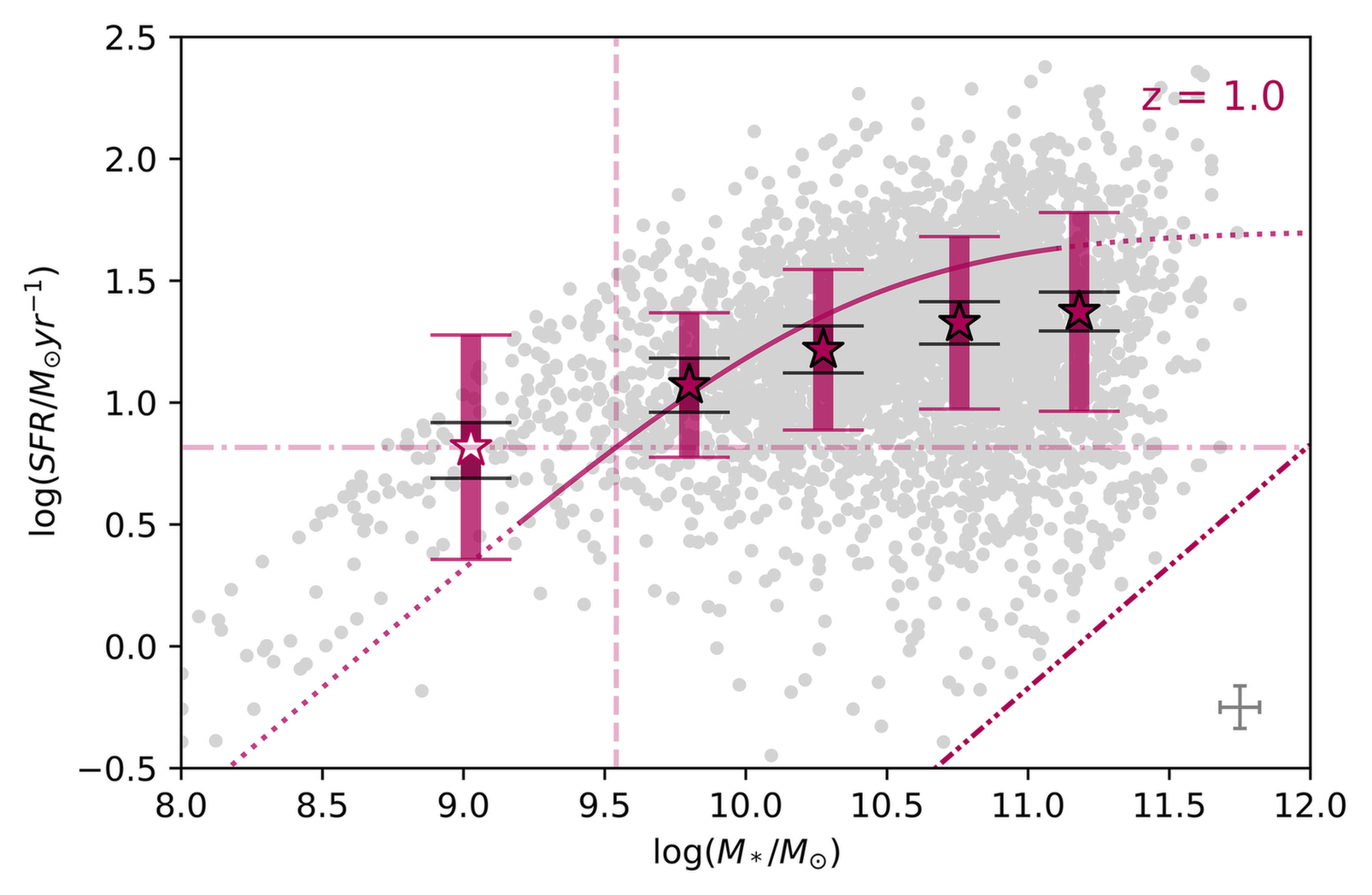} \\ 
    \includegraphics[width=0.48\textwidth]{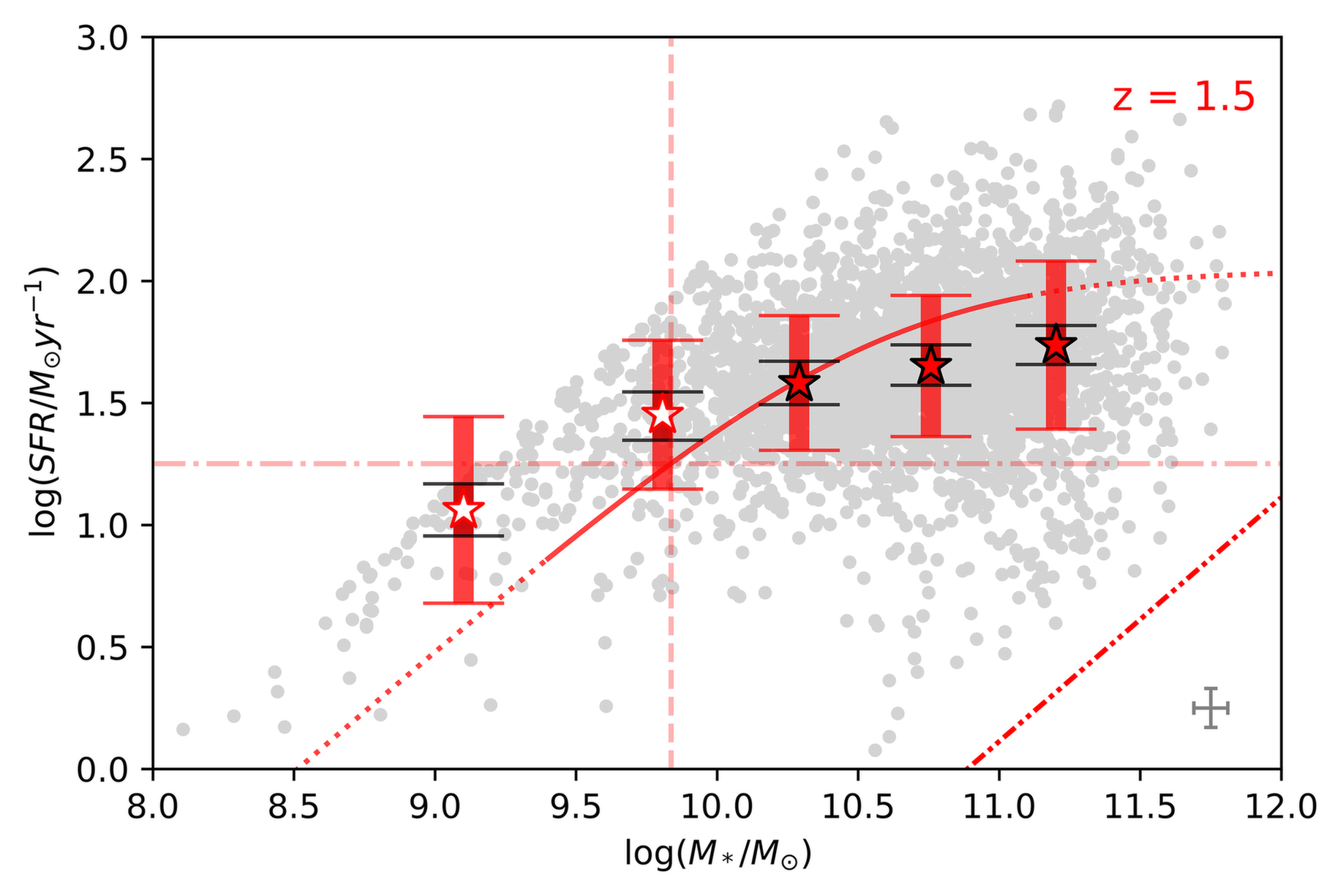} &
    \includegraphics[width=0.48\textwidth]{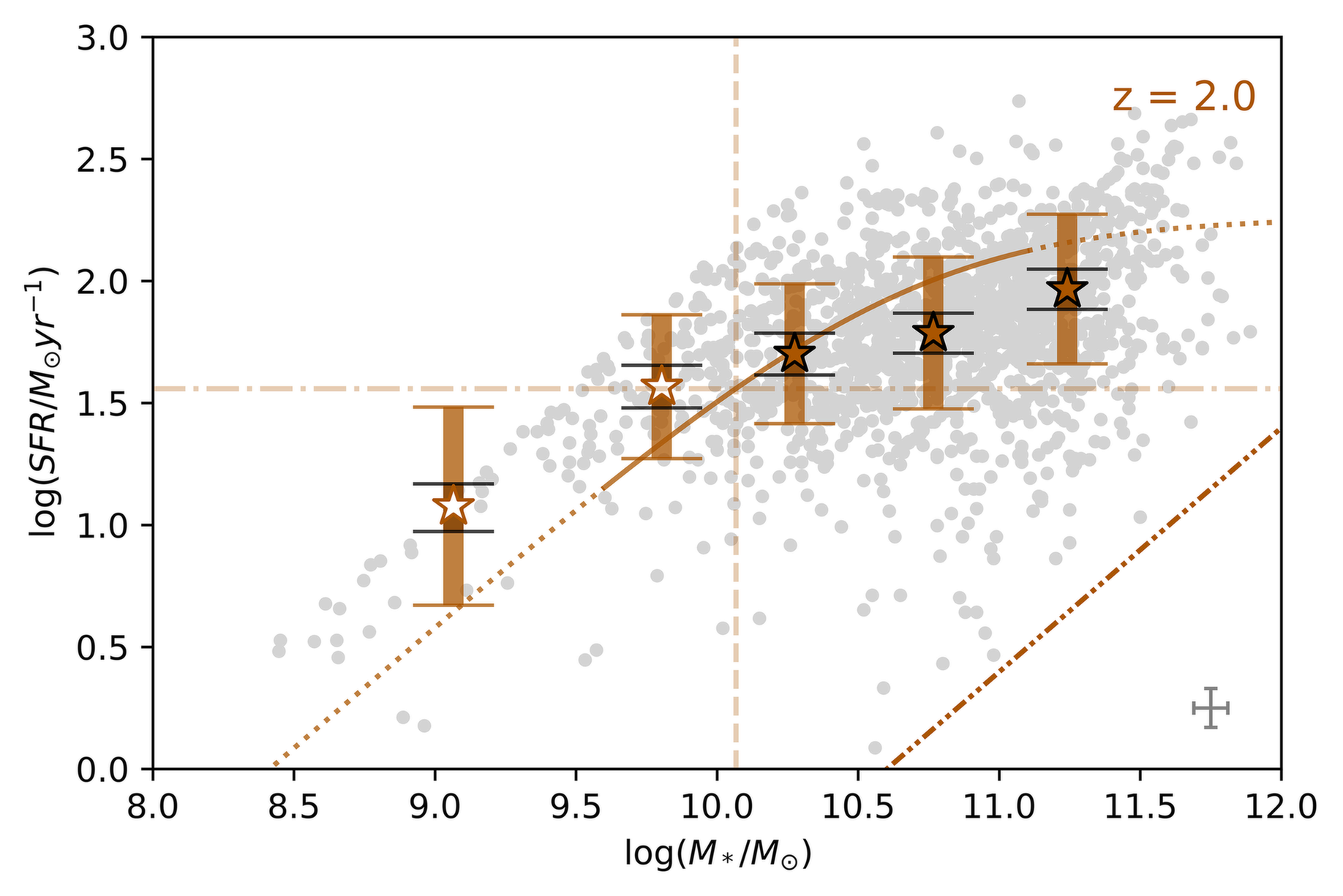} \\
    \includegraphics[width=0.48\textwidth]{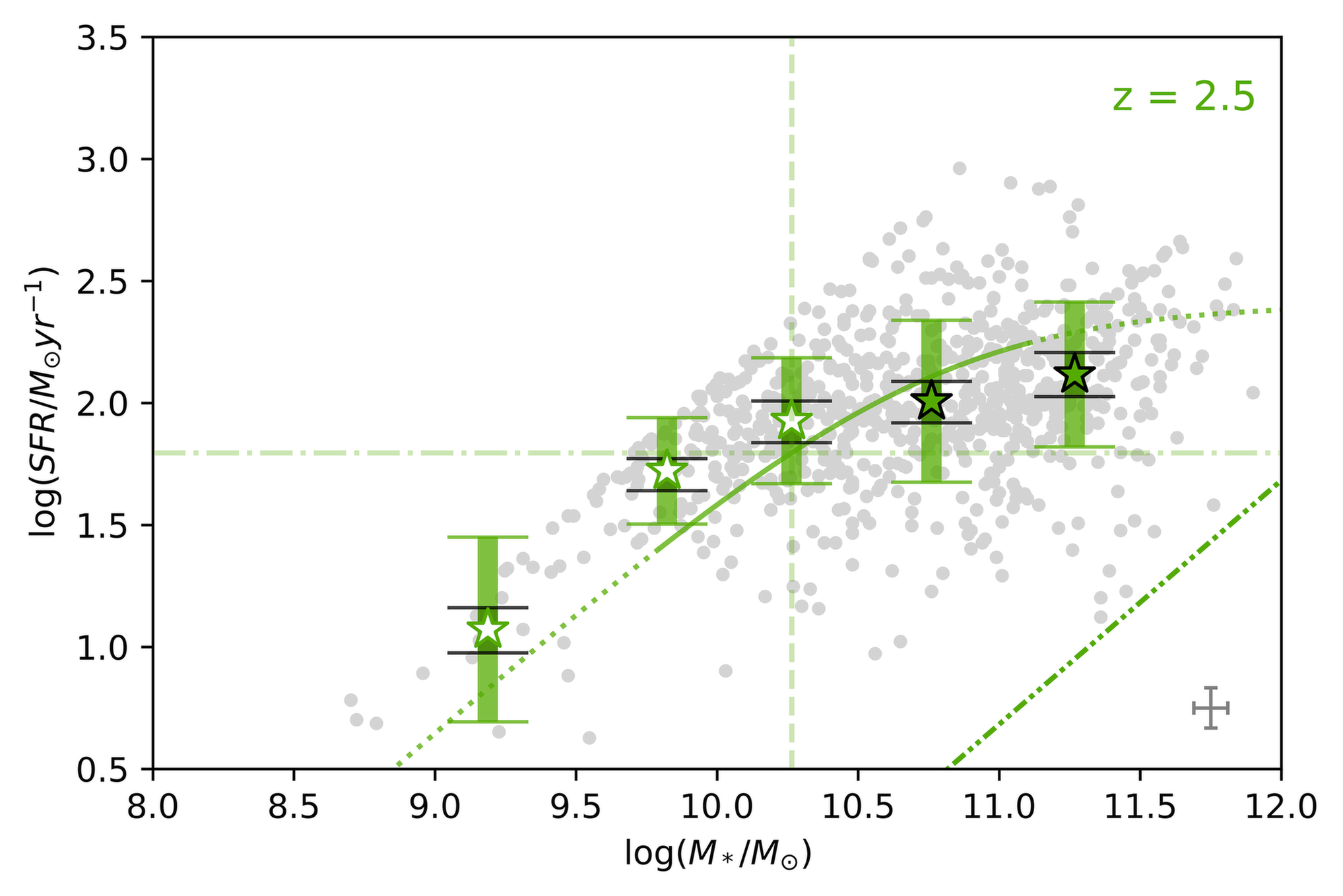} & 
    \includegraphics[width=0.48\textwidth]{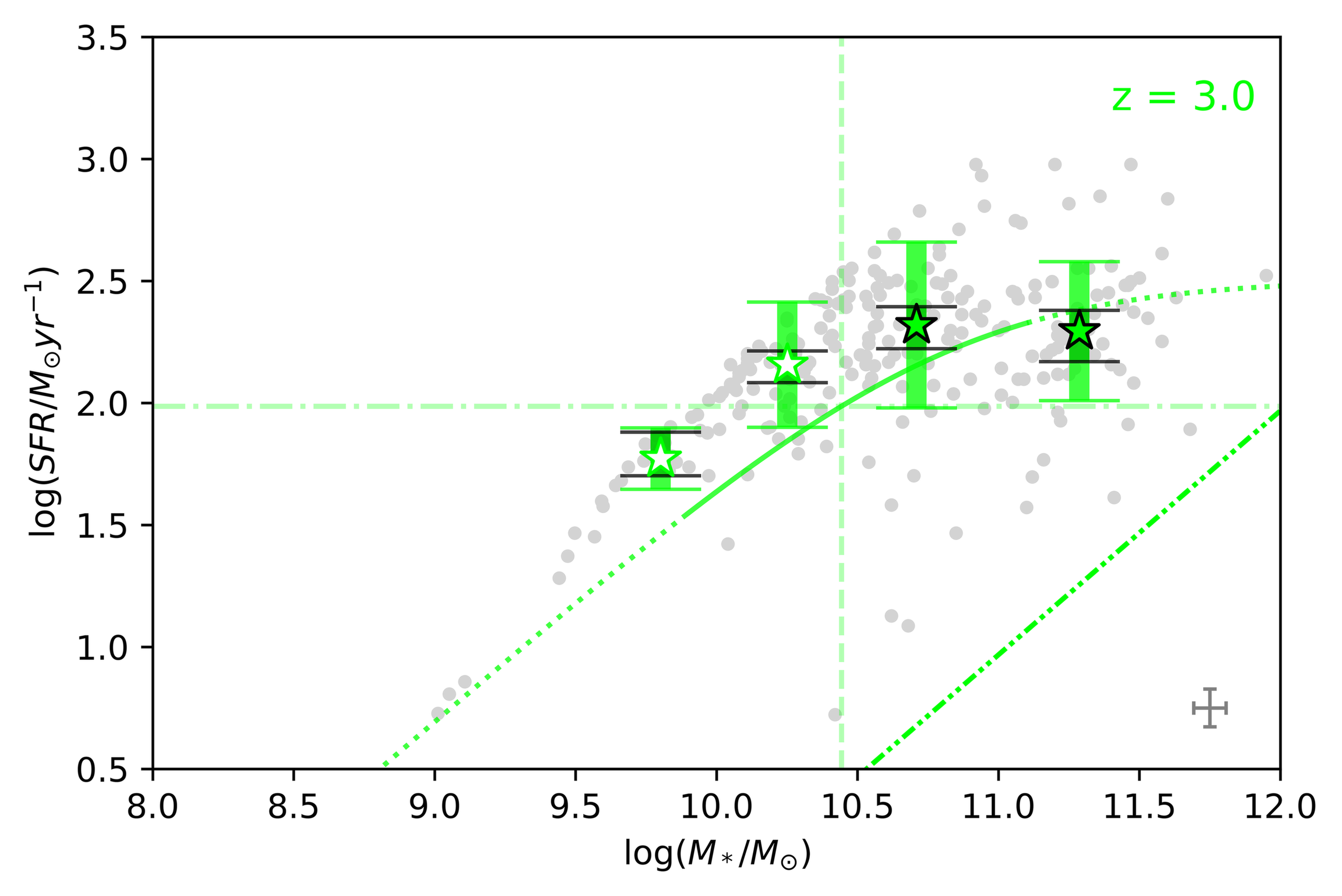} \\
    \end{array}
    $$
    \caption{$M_*$-SFR relation of our galaxies in 6 redshift bins. All bins are shown through 13,071 selection galaxies with colours from purple to red to green coding the range of $z_{{\rm phot}}$ at 0.5, 1.0, 1.5, 2.0, 2.5, and 3.0. The horizontal dot-dashed line indicates the IR-selection completeness cut in SFR, and the vertical dashed line indicates the corresponding IR-selection completeness cut in $M_*$. The open symbols indicate the mass-incomplete data. We drop the data points where the numbers of galaxies within a bin are less than 50. The longer error bars indicate the standard deviation of the SFR distribution in each stellar mass bin, while the shorter black ones represent the median MAGPHYS measurement uncertainties. The gray error bar in the bottom right of each panel denotes the median uncertainties on $M_*$ (along the x-axis) and SFR (along the y-axis) from \texttt{MAGPHYS+photo-z} for the entire redshift bin. The curves are the MS relations at each redshift epoch from \citet{leslie20}. The maximum value of $\log({\rm sSFR}/{\rm yr}^{-1})\sim-8$ due to the adopted SFR timescale shows as an upper boundary of the slope in each SFR-$M_*$ panel, while the bottom right slope represents the sSFR cut at each redshift. }
    \label{fig: 6 panels}
\end{figure*}

For each bin, we assume that any excess in the SFR dispersion relative to the median measurement uncertainty in SFR from \texttt{MAGPHYS+photo-z} is due to the intrinsic scatter of the MS relation. We determine the intrinsic MS scatter by assuming the measured scatter is a result of the measurement uncertainty and intrinsic MS scatter being added in quadrature, which can be rearranged as:  
\begin{center}
\begin{equation}
\begin{split}
&\log(\sigma_{\rm{int}}/M_{\odot}{\rm yr}^{-1}) \\
= &\sqrt{(\log(\sigma_{\rm{tot}}/M_{\odot}{\rm yr}^{-1}))^2-(\log(\sigma_{\rm{meas}}/M_{\odot}{\rm yr}^{-1}))^2},
\label{IS}
\end{split}
\end{equation}
\end{center}
where \textit{$\sigma_{\rm{int}}$}, \textit{$\sigma_{\rm{tot}}$}, and \textit{$\sigma_{\rm{meas}}$} are intrinsic MS scatter of the galaxies, the observed standard deviations, and the median \texttt{MAGPHYS+photo-z} uncertainty, respectively. Figure \ref{fig: 6 panels} displays the galaxy MS for $0.5 \lesssim z_{{\rm phot}} \lesssim 3.0$, showing that the dispersion in SFR of the galaxies are significantly larger than the measurement uncertainties  (representative error bars in lower-right of each panel). In each interval of $M_*$, the size of the SFR intrinsic MS scatter is $0.15 - 0.39$ dex larger than the measurement uncertainty. The horizontal dashed lines in Figure \ref{fig: 6 panels} indicate the limit on SFR defined in Equation \ref{eq: SFR IR-selection} and Figure \ref{fig: IR-selection} at the median $z_{{\rm phot}}$ for each bin. The vertical dashed lines correspond to the stellar mass at this SFR from the reference MS relation. We define the right half in each panel as the mass-complete area. There are 12,380 galaxies (94.71\%) in the mass-complete regions and 691 galaxies (5.29\%) in the mass-incomplete regions. We show the values of the scatter terms for our mass-complete bins in Table \ref{table: Main results of this paper}. 

\begin{figure*}
    \centering
    $$
    \begin{array}{cc}
    \includegraphics[width=0.48\textwidth]{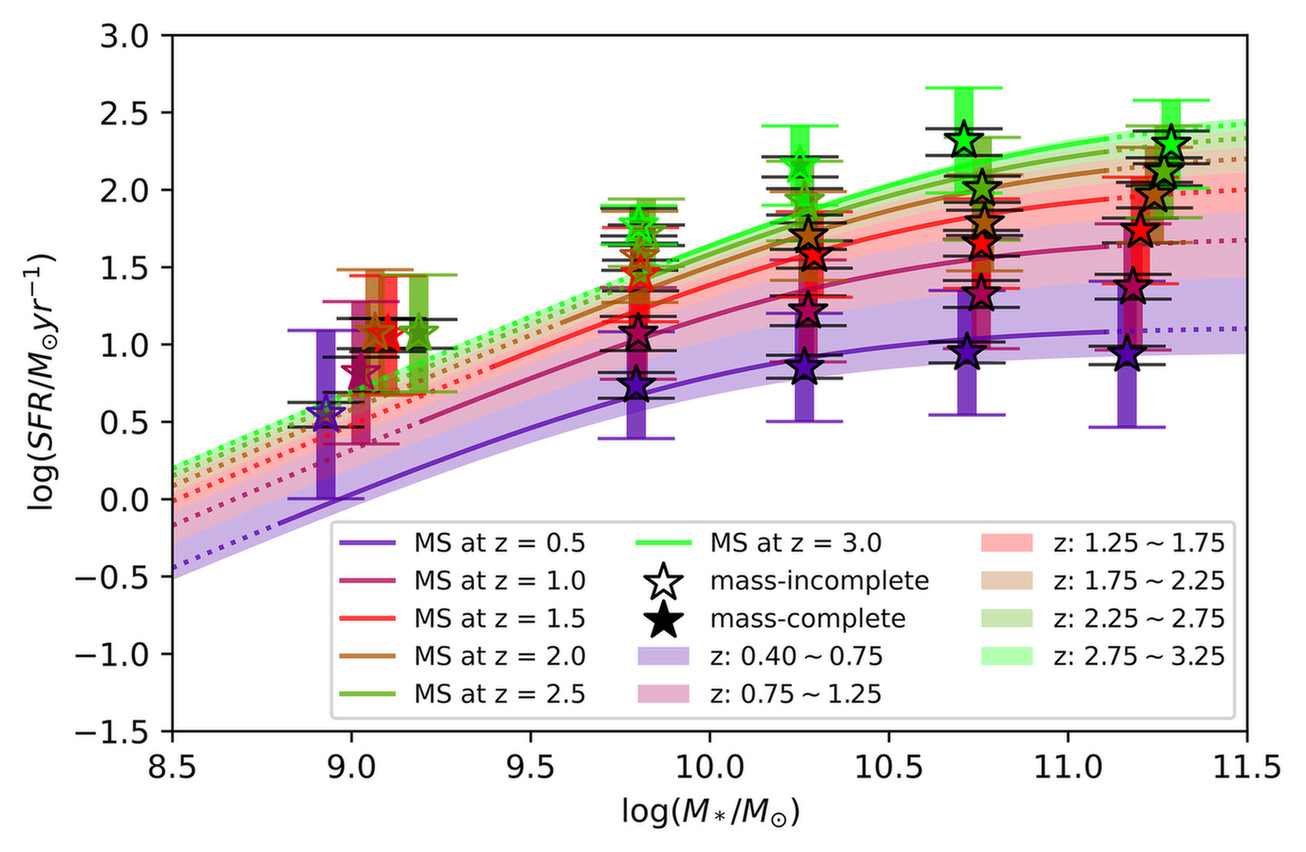} &
    \includegraphics[width=0.48\textwidth]{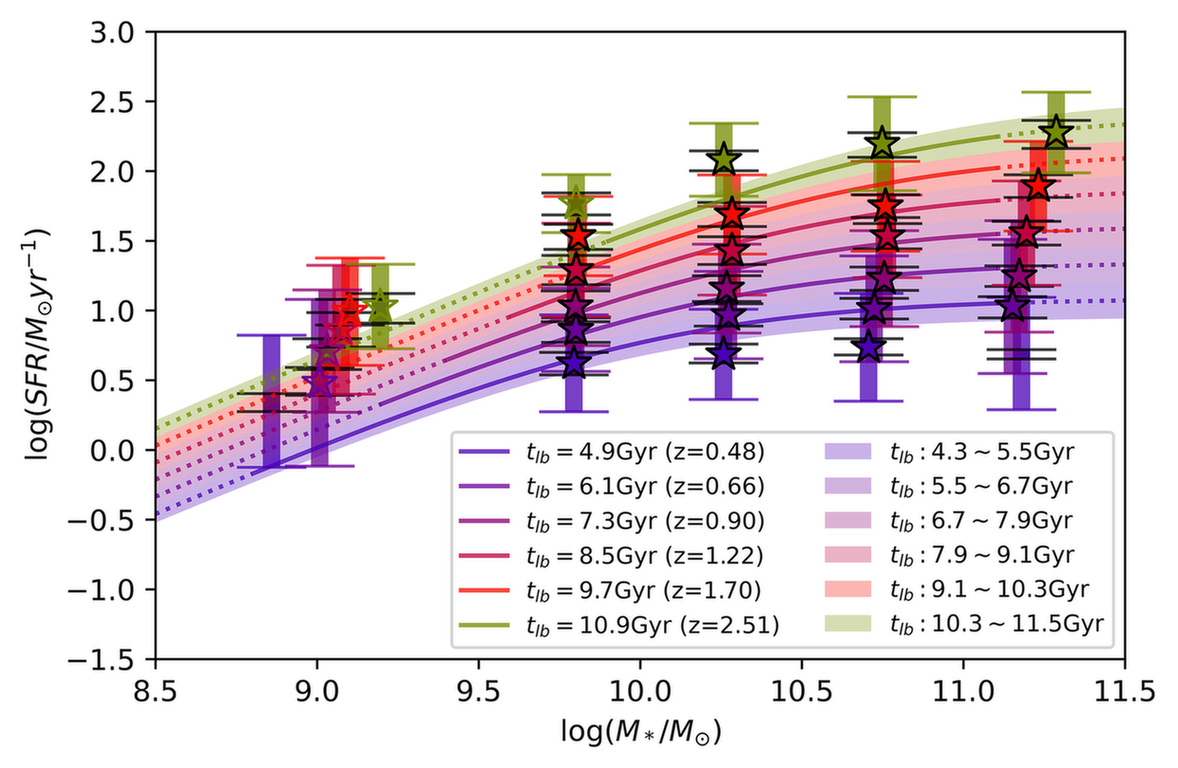} \\
    \end{array}
    $$
    \caption{The MS evolution and scatter for our sample with the same colour scheme adopted in Figure \ref{fig: 6 panels}. The shaded regions shown in the diagrams indicate the range of the reference MS between the upper and lower redshift boundaries for each bin and do not relate to the intrinsic MS scatter. The thick and thin error bars indicate the observed standard deviation in SFR and the median SFR uncertainty from \texttt{MAGPHYS+photo-z} for each bin, respectively. The left panel shows the sample binned in equal redshift bins, with the mass-incomplete bins shown as open symbols. The right panel is similar to the left but binned in equal look-back time bins. Adopting equal-width redshift bins (left-panel) instead of look-back time (right-panel) may impact the measured intrinsic MS scatter because of the evolution in the MS relation over the redshift range contained within a single bin (i.e., differing widths in shaded region in left-panel relative to right-panel). Hence, we adopt look-back time bins for our analysis. }
    \label{fig: MS for all redshifts}
\end{figure*}

We observe the following three trends (see the left panel of Figure \ref{fig: MS for all redshifts} and Table \ref{table: Main results of this paper}). First, the median SFR measurement uncertainties are always smaller than the intrinsic MS scatter of the SFR. The minimum difference between intrinsic MS scatter and uncertainties is $0.13$ dex in the range of $10^{10} - 10^{10.5}M_{\odot}$ at $z_{{\rm phot}} = 2.5$, while the maximum occurring at $z_{{\rm phot}} = 0.5$ for galaxies with $\log(M_*/M_{\odot}) > 11.0$ is $0.39$ dex. Second, excluding the mass-incomplete regions, our galaxies roughly follow the same observed main sequence as shown in \citet{leslie20} (i.e. Equation \ref{eq: SFR IR-selection}), but with slightly lower SFRs than the reference MS for most redshift bins. Third, the intrinsic dispersion in SFR at a given mass tends to decrease as the $z_{{\rm phot}}$ increases at a fixed $M_*$. 

The size of the $z_{{\rm phot}}$ interval we selected may affect the behaviour of the SFR intrinsic MS scatter evolution. The width of each $z_{{\rm phot}}$ in our criteria is $\Delta z_{{\rm phot}} = 0.5$ except $0.35$ at $z_{{\rm phot}} = 0.5$. However, with the increase of $z_{{\rm phot}}$, the cosmic time corresponding to the $\Delta z_{{\rm phot}}$ is decreasing because the look-back time $t_{{\rm lb}}$ does not linearly increase with $z_{{\rm phot}}$. As a result, this reducing length of the binning interval in cosmic time with increasing $z_{{\rm phot}}$ may affect the SFR intrinsic MS scatter. To examine this issue, we adopt look-back time ($t_{{\rm lb}}$) instead of redshift as a more consistent way to measure the intrinsic MS scatter. We convert the $z_{{\rm phot}}$ into $t_{{\rm lb}}$ and rearrange our sample of 13,071 galaxies in 6 $t_{{\rm lb}}$ bins (4.9, 6.1, 7.3, 8.5, 9.7, 10.9Gyr) with the equal length of time ($\Delta t_{{\rm lb}} = 1.2$Gyr). We reproduce the $\log({\rm SFR})$-$\log(M_*)$ plane in the right panel of Figure \ref{fig: MS for all redshifts} and present the results in Table \ref{table: Main results of this paper in t_lb}. We find that the $t_{{\rm lb}}$ results share the same trends and features as the previous $z_{{\rm phot}}$ version. However, because binning the data in equal $t_{{\rm lb}}$ width removes the unequal-length effect when measuring the intrinsic MS scatter, we will adopt this for our main analysis.

\begin{figure*}
     \centering
    $$
    \begin{array}{cc}
    \centering
    \includegraphics[width=0.48\textwidth]{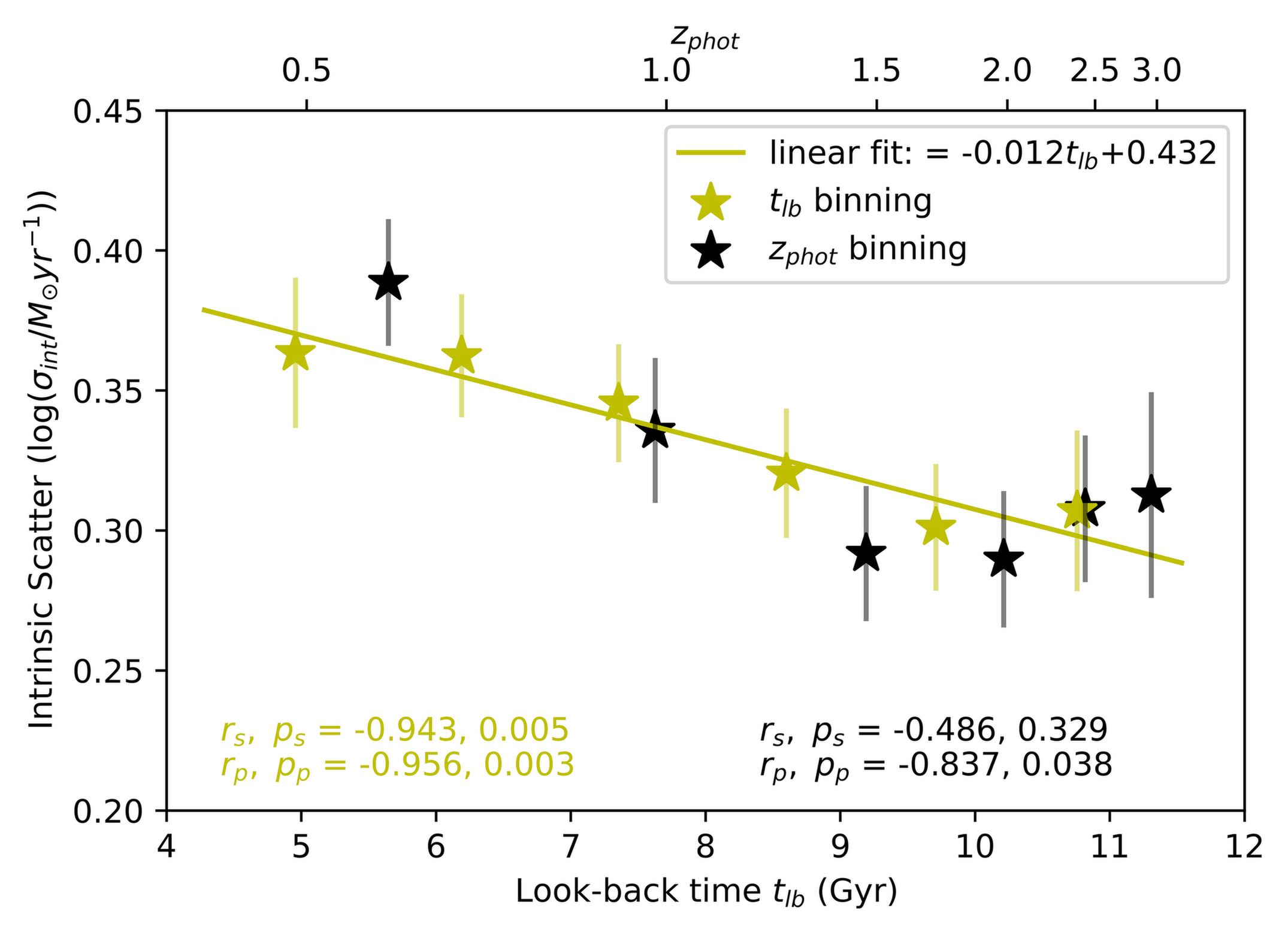} &
    \includegraphics[width=0.48\textwidth]{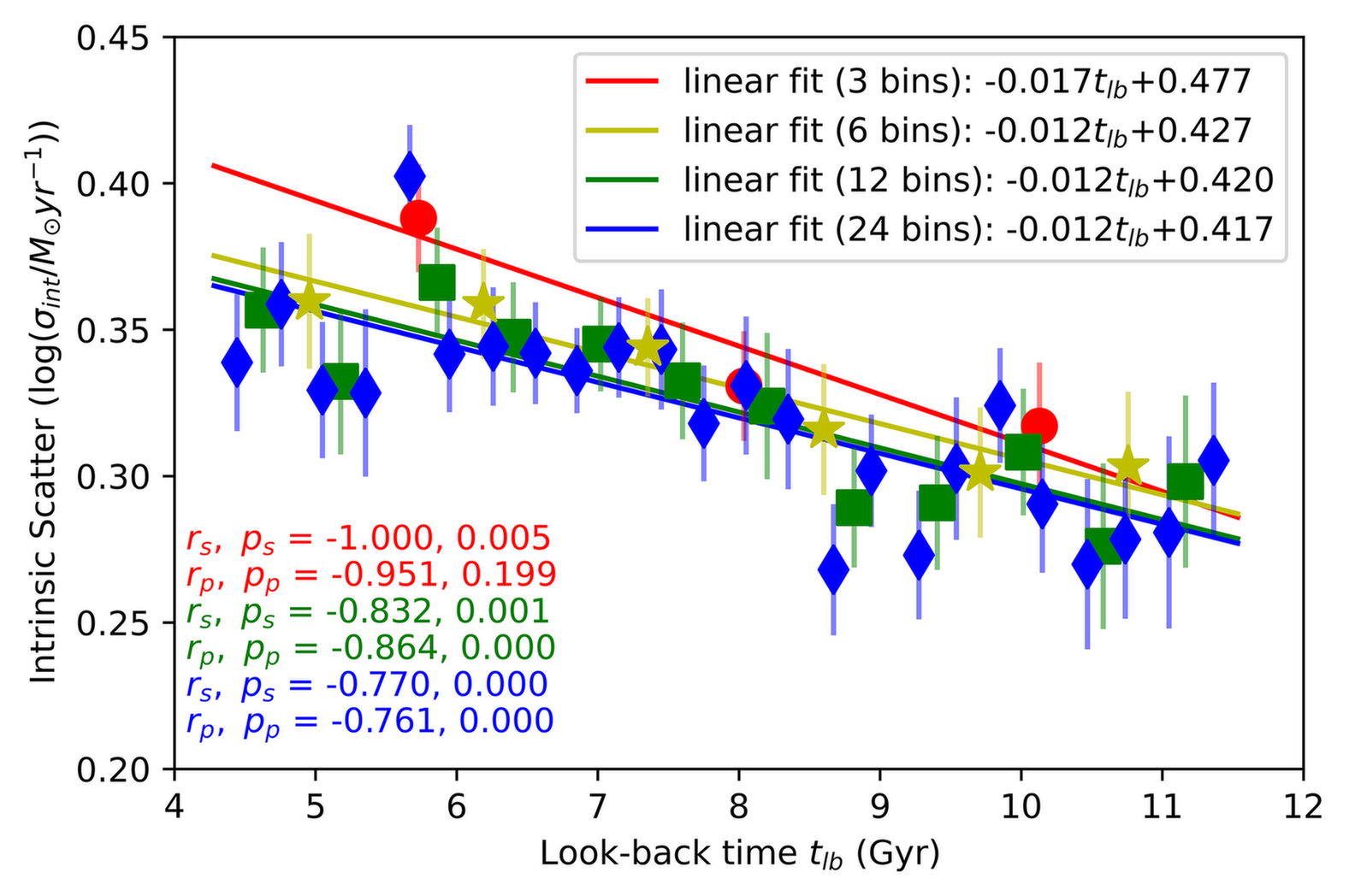} \\ 
    \end{array}
    $$
    \caption{The left panel is intrinsic MS scatter vs $z_{{\rm phot}}$ and $t_{\rm lb}$ binning with a linear regression fit to the look-back time bins. The yellow star-like scatter points are the $t_{{\rm lb}}$ binning data. We take the average of the data points in each mass bin and obtain error bars of intrinsic MS scatter by bootstrap resampling the distribution 100 times based on varying the individual values by their uncertainties and rebinning them. The solid yellow line is the linear fit of intrinsic MS scatter versus $t_{{\rm lb}}$. The right panel display the effect of binning number. A total of 12,380 galaxies are regrouped in 3, 6, 12 and 24 bins with red circles, yellow star, green squares and blue diamonds. In each panel, we show the Spearman and Pearson correlation coefficients ($r_s$ and $r_p$) as well as the corresponding p-values ($p_s$, $p_p$) for different binning data. }       
    \label{fig: IS vs z_phot}
\end{figure*}

The left panel of Figure \ref{fig: IS vs z_phot} shows the relationship between intrinsic MS scatter and $t_{{\rm lb}}$ for both the redshift and look-back time binning. In this study, we adopt the weighted linear regression to the intrinsic MS scatter versus $t_{{\rm lb}}$:
\begin{center}
\begin{equation}
\begin{split}
\log(\sigma_{\rm{int}}/M_{\odot}{\rm yr}^{-1}) = (-0.012\pm0.002)t_{{\rm lb}}+(0.432\pm0.015), \\
\label{eq: IS vs z_phot}
\end{split}
\end{equation}
\end{center}
where $\sigma_{\rm{int}}$ is the intrinsic scatter of the MS, and these parameters are calculated with mass-complete sample. We observe a trend of decreasing intrinsic MS scatter up to $9.7$Gyr ($z \sim 1.7$) as $t_{{\rm lb}}$ increases. The error bars are derived by bootstrap resampling the data in each bin 100 times. Bins with smaller sample sizes have larger bootstrap errors. Although the descending rate of intrinsic MS scatter over look-back time is shallow, the Spearman and Pearson correlation coefficients ($r_s = -0.943$ and $r_p = -0.956$, respectively) indicate a strong monotonic decreasing correlation. Conversely, with $r_s = -0.486$ and $r_p = -0.837$, there is a weaker and tentative correlation when using redshift binning. This is due to the redshift binning having a potential upturning feature at $z_{\rm phot} \geq 2$. We suggest this is a consequence of unequal-length binning for redshift, which will be discussed in the next paragraph. Furthermore, there is a `upturn' feature, and the intrinsic MS scatter tends to increase after $t_{{\rm lb}} \sim 10 {\rm Gyr}$. Given the uncertainty in our intrinsic MS scatter and our limited sample size at high-$z$, it is difficult to assess the significance of this upturn with our current data.     

The intrinsic MS scatter may vary with the adopted $\Delta t_{{\rm lb}}$ of each bin. The right panel of Figure \ref{fig: IS vs z_phot} presents the effect of binning the data in different $\Delta t_{{\rm lb}}$ widths. The data are binned into 3, 6, 12, and 24 groups (no less than 100 galaxies in each bin) in 4 different sets with equal $t_{{\rm lb}}$ widths, where 6 is our fiducial number of bins. It can be seen that the data in the 12 and 24 bins have a similar distribution statistically relative to our fiducial binning. We find that as the number of bins increases, the normalisation (intrinsic MS scatter) slightly decreases. However, the coefficients of the corresponding equations tend to converge somewhere closely below the current linear regression equation (i.e. the yellow line). Even though the decreasing binning time scale causes more severe fluctuation along the linear regression line, the similarity and high absolute values of $r_s$ and $r_p$ still demonstrate a strong correlation between intrinsic MS scatter and $t_{{\rm lb}}$. On the other hand, this phenomenon also partially explains why redshift binning is not a good approach in this study, especially at high redshifts: a narrower binning time scale may lead to larger fluctuations in the intrinsic MS scatter. Since there is no significant discrepancy in intrinsic MS scatter for $n_{\rm{bin}} \geq 6$, we expect that these coefficients in linear regression lines approach some values slightly smaller than the current binning one. Hence, we conclude that the binning does not strongly affect the trend of intrinsic MS scatter evolution and we adopt $n_{\rm{bin}} \geq 6$ as the current $t_{{\rm lb}}$ binning number. 

\begin{table*}
\centering
\begin{tabular}{|c|cccc|cccc|cccc|cccc|}
\hline
$M_*$ & \multicolumn{4}{c|}{$9.5-10.0\ \rm Log(M_{\odot})$} & \multicolumn{4}{c|}{$10.0-10.5\ \rm Log(M_{\odot}$)} & \multicolumn{4}{c|}{$10.5-11.0\ \rm Log(M_{\odot})$} & \multicolumn{4}{c|}{$>11.0\ \rm Log(M_{\odot}$)} \\
\hline
$z_{{\rm phot}}$ & N & $\sigma_{\rm{tot}}$ & $\sigma_{\rm{meas}}$ & $\sigma_{\rm{int}}$ & N & $\sigma_{\rm{tot}}$ & $\sigma_{\rm{meas}}$ & $\sigma_{\rm{int}}$ & N & $\sigma_{\rm{tot}}$ & $\sigma_{\rm{meas}}$ & $\sigma_{\rm{int}}$ & N & $\sigma_{\rm{tot}}$ & $\sigma_{\rm{meas}}$ & $\sigma_{\rm{int}}$ \\
\cline{1-17}
$0.5$ & 428 & 0.36 & 0.09 & 0.34 & 897 & 0.35 & 0.08 & 0.34 & 922 & 0.40 & 0.07 & 0.40 & 239 & 0.47 & 0.06 & 0.47 \\
$1.0$ & 365 & 0.30 & 0.11 & 0.28 & 1036 & 0.34 & 0.10 & 0.32 & 1660 & 0.35 & 0.09 & 0.34 & 831 & 0.41 & 0.08 & 0.40 \\
$1.5$ & 247 & 0.31 & 0.10 & 0.29 & 776 & 0.28 & 0.09 & 0.26 & 1430 & 0.29 & 0.08 & 0.28 & 953 & 0.34 & 0.08 & 0.34 \\
$2.0$ & 102 & 0.29 & 0.09 & 0.28 & 300 & 0.29 & 0.09 & 0.27 & 663 & 0.31 & 0.08 & 0.30 & 616 & 0.31 & 0.08 & 0.29 \\
$2.5$ & 69 & - & - & - & 202 & 0.26 & 0.09 & 0.24 & 227 & 0.35 & 0.09 & 0.33 & 261 & 0.30 & 0.09 & 0.28 \\
$3.0$ & 25 & - & - & - & 78 & - & - & - & 74 & 0.37 & 0.09 & 0.36 & 72 & 0.28 & 0.10 & 0.26 \\
\hline
\end{tabular}
\caption{"N" is the number of galaxies, "$\sigma_{\rm{tot}}$" is the galaxy SFR dispersion (standard deviation), "$\sigma_{\rm{meas}}$" is the MAGPHYS uncertainty in SFR and "$\sigma_{\rm{int}}$" is the intrinsic MS scatter in each $M_*$ bin. The mass-incomplete data are marked as "-", but the number of galaxies in the binning interval are still shown. Since all the $M_*<9.5\log(M_\odot)$ galaxies lie in the mass-incomplete regime, they are not included in the table. }
\label{table: Main results of this paper}
\end{table*}

\begin{table*}
\centering
\begin{tabular}{|c|cccc|cccc|cccc|cccc|}
\hline
$M_*$ & \multicolumn{4}{c|}{$9.5-10.0\ \rm Log(M_{\odot}$)} & \multicolumn{4}{c|}{$10.0-10.5\ \rm Log(M_{\odot}$)} & \multicolumn{4}{c|}{$10.5-11.0\ \rm Log(M_{\odot})$} & \multicolumn{4}{c|}{$>11.0\ \rm Log(M_{\odot}$)} \\
\hline
$t_{{\rm lb}}$(Gyr) & N & $\sigma_{\rm{tot}}$ & $\sigma_{\rm{meas}}$ & $\sigma_{\rm{int}}$ & N & $\sigma_{\rm{tot}}$ & $\sigma_{\rm{meas}}$ & $\sigma_{\rm{int}}$ & N & $\sigma_{\rm{tot}}$ & $\sigma_{\rm{meas}}$ & $\sigma_{\rm{int}}$ & N & $\sigma_{\rm{tot}}$ & $\sigma_{\rm{meas}}$ & $\sigma_{\rm{int}}$ \\
\cline{1-17}
4.9(z=0.48) & 226 & 0.36 & 0.08 & 0.35 & 427 & 0.32 & 0.07 & 0.32 & 343 & 0.39 & 0.06 & 0.39 & 40 & - & - & - \\
6.1(z=0.66) & 222 & 0.31 & 0.09 & 0.29 & 532 & 0.32 & 0.08 & 0.31 & 674 & 0.38 & 0.07 & 0.37 & 195 & 0.48 & 0.08 & 0.47 \\
7.3(z=0.90) & 240 & 0.28 & 0.11 & 0.26 & 648 & 0.32 & 0.09 & 0.31 & 1082 & 0.35 & 0.09 & 0.33 & 504 & 0.40 & 0.07 & 0.39 \\
8.5(z=1.22) & 219 & 0.34 & 0.11 & 0.32 & 696 & 0.33 & 0.10 & 0.31 & 1128 & 0.30 & 0.09 & 0.28 & 758 & 0.38 & 0.09 & 0.37 \\
9.7(z=1.70) & 280 & 0.28 & 0.09 & 0.27 & 855 & 0.29 & 0.09 & 0.27 & 1607 & 0.32 & 0.08 & 0.31 & 1316 & 0.32 & 0.08 & 0.31 \\
10.9(z=2.51) & 49 & - & - & - & 131 & 0.26 & 0.07 & 0.25 & 140 & 0.35 & 0.09 & 0.34 & 157 & 0.29 & 0.10 & 0.27 \\ 
\hline
\end{tabular}
\caption{Look-back time version of Table \ref{table: Main results of this paper} by collecting and reassigning the 12,380 mass-complete data to 6 $t_{{\rm lb}}$ bins. We exclude bins in which the number of galaxies is less than 50 ($>10^{11}M_{\odot}$ at $t_{{\rm lb}} = 4.9$Gyr), together with the mass-incomplete data. In general, the intrinsic MS scatter is still significantly larger than measurement uncertainty in new binning criteria, which avoids the effect of nonlinear $t_{{\rm lb}}$ widths in previous $z_{{\rm phot}}$ binning. }
\label{table: Main results of this paper in t_lb}
\end{table*}

\section{Discussion}
\label{Discussion}
 
\subsection{Comparison to the halo mass-stellar mass relation}
\label{Comparison to the halo mass-stellar mass relation}

\begin{figure*}
     \centering
    $$
    \begin{array}{cc}
    \centering
    \includegraphics[width=0.48\textwidth]{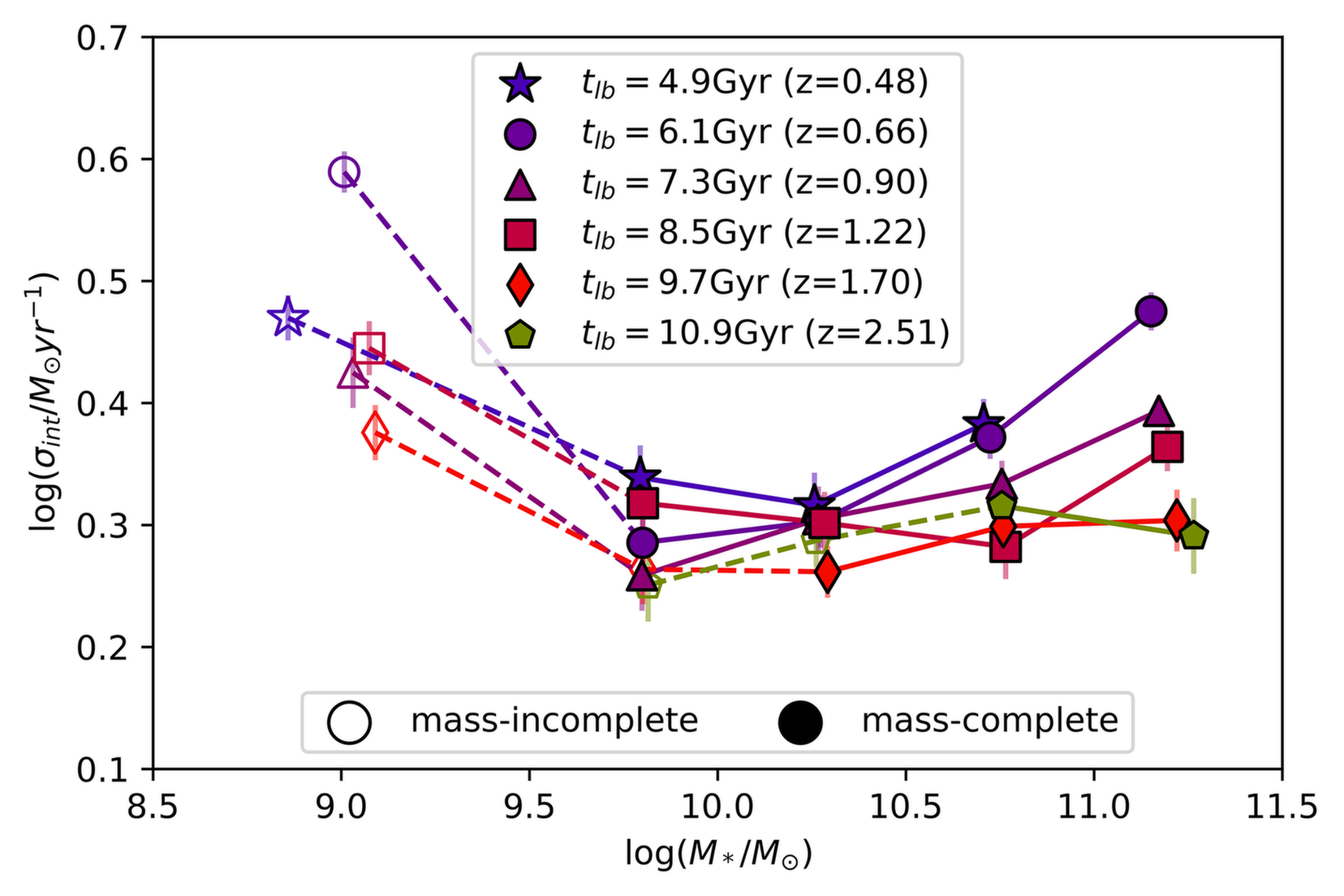} &
    \includegraphics[width=0.48\textwidth]{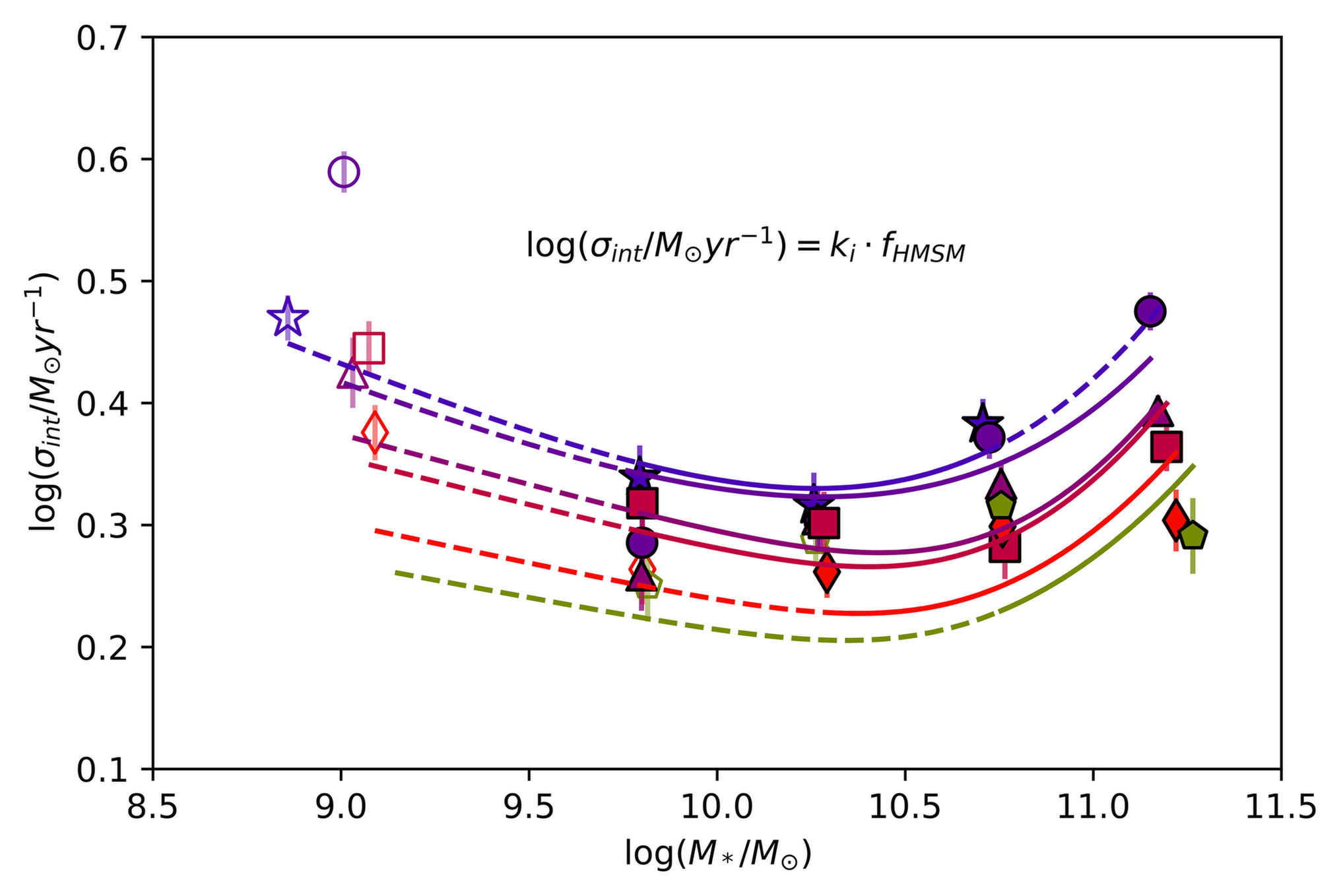} \\ 
    \end{array}
    $$
    \caption{Left panel: the intrinsic MS scatter $\sigma(\log({\rm SFR}))$ vs $M_*$ for our 6 look-back time bins. Solid lines connect the mass-complete data and dashed lines indicate the mass-incomplete regions. Right panel: we show a fit of our toy model (relation indicated in the panel), based on the halo mass to stellar mass fraction vs stellar mass relation \citet{behroozi10} normalized by a constant factor, normalized by a constant factor, relative to our intrinsic MS scatter. Similar to the left, we indicate the mass-complete and mass-incomplete regions with solid and dashed lines, respectively. }      
    \label{fig: IS vs mass}
\end{figure*}

The stochastic events that occurred throughout an SF galaxy's history, including as galaxy mergers and supernova \& AGN feedback, are assumed to be responsible for the inherent MS scatter. The amount of burstiness in SFH, which is probably related to galactic feedback mechanisms, is reflected in the distribution and evolution of intrinsic MS scatter.

In the left panel of Figure \ref{fig: IS vs mass}, we show the distribution of the intrinsic MS scatter versus $M_*$. We see a minimum intrinsic MS scatter of $\sim0.35$ dex at $M_* \sim 10^{10.25}M_{\odot}$. We see a higher increase of $\sim 0.1$ dex at higher mass ($> 10^{11}M_{\odot}$) with decreasing look-back time for low redshifts ($t_{\rm lb}$: 4.9 - 8.5 Gyr) and a relatively flat relation at higher redshifts ($t_{\rm lb} \gtrsim 9.7 {\rm Gyr}$). In the low mass end ($\sim 10^{9}M_{\odot}$), where the galaxies are mass-incomplete, the intrinsic scatter rises from 0.35 to 0.6 dex. For some of the redshift bins, this type of trend is qualitatively similar to the turnover that occurs in the halo mass-stellar mass (HMSM) relation (see Figure 2 of \citet{wechsler18}). The shape of the HMSM relation is thought to be a consequence of feedback, with SF feedback reducing the SF efficiency in low-mass galaxies and AGN feedback reducing the SF efficiency in high-mass galaxies, with a turnover at halo mass $M_{\rm{h}} \sim 10^{12}M_{\odot}$, which is coincidentally corresponding to the upturn point of $M_*$-$\sigma_{\rm int}$ panel at $M_* \sim 10^{10.25}M_{\odot}$ in this study. The intrinsic scatter in the MS is also expected to be linked to feedback, and this may account for similarities in the observed trends with the HMSM relation. 

In the right panel of Figure \ref{fig: IS vs mass}, we present a toy model where we relate the HMSM relation with the intrinsic MS scatter. We adopt the best-fit functional form and the parameterised data of the stellar mass halo mass (SMHM) relation from Equations 21 \& 22 and Table 2 of \citet{behroozi10}. \citet{behroozi10} parameterize the evolution of SMHM relation in terms of $M_1$, $M_{*,0}$, $\beta$, $\delta$ and $\gamma$. All these variables vary with the scale factor $a$. For our model, we convert the standard SMHM relation into HMSM fraction vs stellar mass (i.e., $\log(M_{\rm{h}}/M_*)$ vs $\log(M_*)$, instead of $\log(M_*/M_{{\rm h}})$ vs $\log(M_{\rm{h}})$. This change results in the HMSM relation having an upturn shape instead of the usual downturn shape (because we invert the values in the y-axis ratio). This modified HMSM relation presents a similar turnover feature as the `U-shaped' distribution shown in the left subplot of Figure \ref{fig: IS vs mass}. Considering time evolution, we renormalize the HMSM relation to match our data by multiplying the HMSM fraction by an arbitrary coefficient $k_i$ ($i$ indicates different $t_{{\rm lb}}$), which is computed by the non-linear regression: 
\begin{center}
\begin{equation}
\begin{split}
\log(\sigma_{\rm{int}}/M_{\odot}{\rm yr}^{-1}) =\ &k_{i}\cdot f_{\rm{HMSM}}(a), \\
\label{eq: IS = k_i*f_HMSM}
\end{split}
\end{equation}
\end{center}
where $\sigma_{\rm{int}}$ is the intrinsic MS scatter, $k_i$ is a constant factor (see Table \ref{table: k_i}) and $f_{\rm{HMSM}}(a) = M_{\rm{h}}/M_*$ is the HMSM fraction that varies with the scale factor, $a$ \citep{behroozi10}. The Equation \ref{eq: IS = k_i*f_HMSM} does a reasonable job of matching the trends for the first four time bins (4.9-8.5Gyr), but the final two bins (9.7 and 10.9Gyr) favor a flatter shape than our toy model at high $M_*$. At the lower and higher mass ends, the increasing intrinsic MS scatter may be driven by the feedback of supernovas and AGNs, respectively. In contrast, in the mid-range of stellar mass ($10.0 <\log(M_*/M_{\odot})< 10.5$), the intrinsic MS scatter relation reaches a minimum (maximum in HMSM relation), reflecting the maximum conversion efficiency of gas to baryon and is thought to be linked to a minimum in the influence of starburst and galaxy feedback. We notice a large discrepancy between our data with shifted HMSM fraction at higher redshifts, which may be due to the low quality of observational data at high redshifts or differences in feedback mechanisms in the earlier universe. For example, high-$z$ observational limitation can lead to larger uncertainties on SFR and make it more difficult to constrain the intrinsic MS scatter. On the other hand, weaker AGN feedback for high-$M_*$ galaxies at high-$z$ may also account for the almost constant intrinsic MS scatter. 

\begin{table}
\centering
\begin{tabular}{|c|c|}
\hline
$t_{{\rm lb}}$ & $k_i$ \\
\hline
4.9Gyr (z=0.48) & 0.20 \\
6.1Gyr (z=0.66) & 0.19 \\
7.3Gyr (z=0.90) & 0.17 \\
8.5Gyr (z=1.22) & 0.15 \\
9.7Gyr (z=1.70) & 0.12 \\
10.9Gyr (z=2.51) & 0.10 \\
\hline
\end{tabular}
\caption{Best-fit values for the constant factor $k_i$ at different bins of look-back time for our toy model given by Equation \ref{eq: IS = k_i*f_HMSM}. }
\label{table: k_i}
\end{table}

\subsection{Comparison to observational studies}
\label{Comparison to observational studies}

\begin{figure*}
     \centering
    $$
    \begin{array}{cc}
    \centering
    \includegraphics[width=0.48\textwidth]{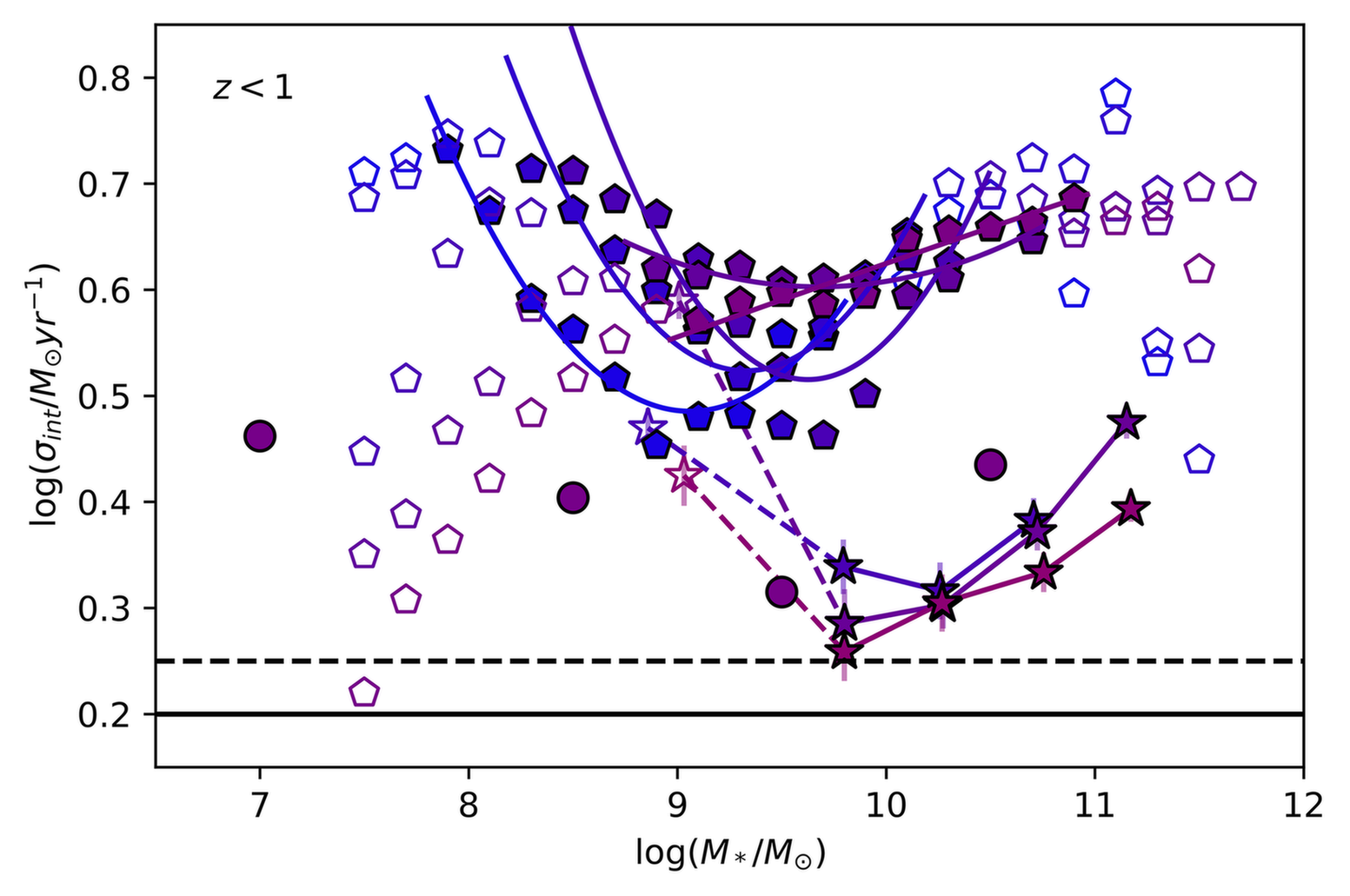} &
    \includegraphics[width=0.48\textwidth]{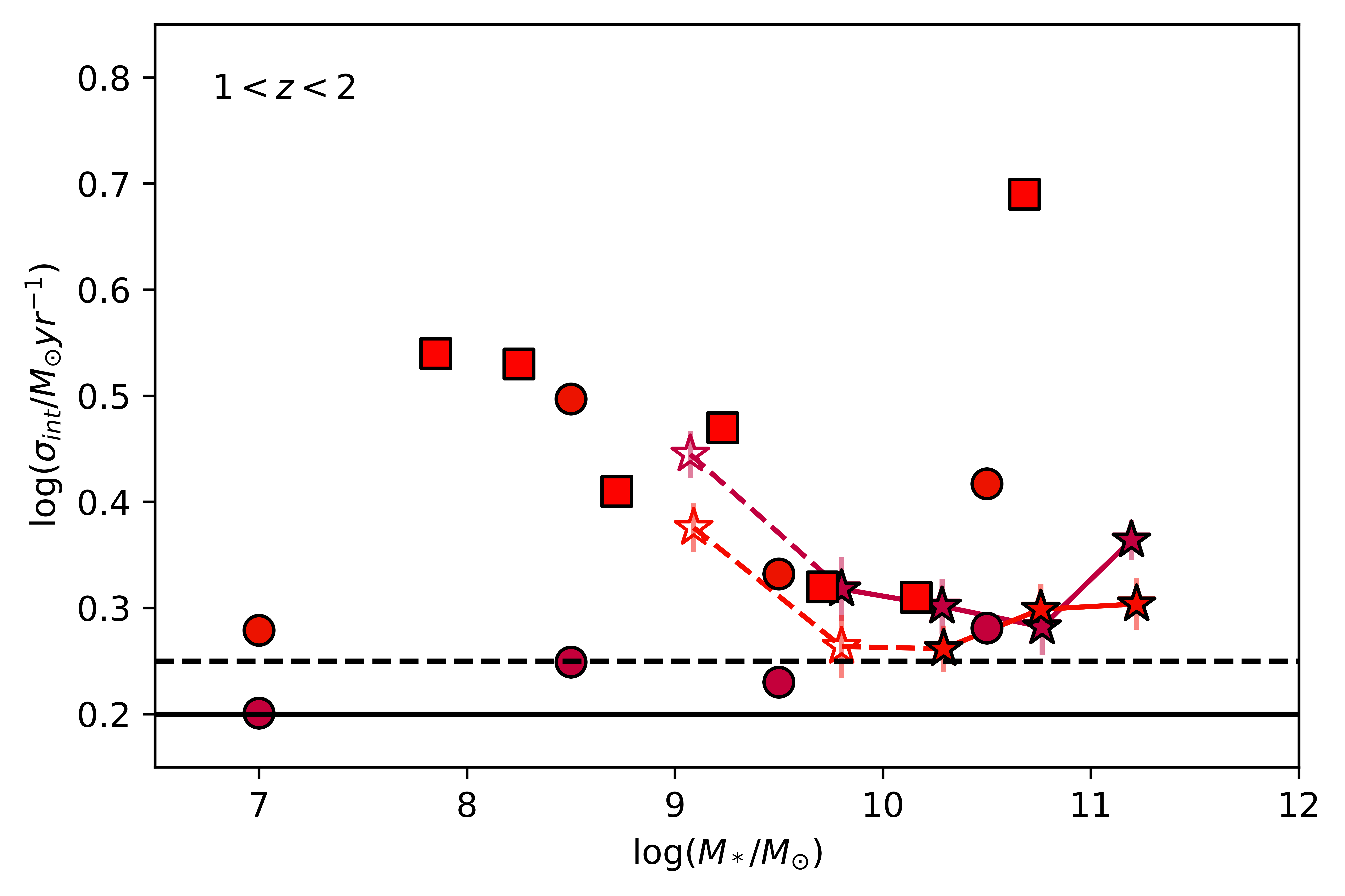} \\ 
    \includegraphics[width=0.48\textwidth]{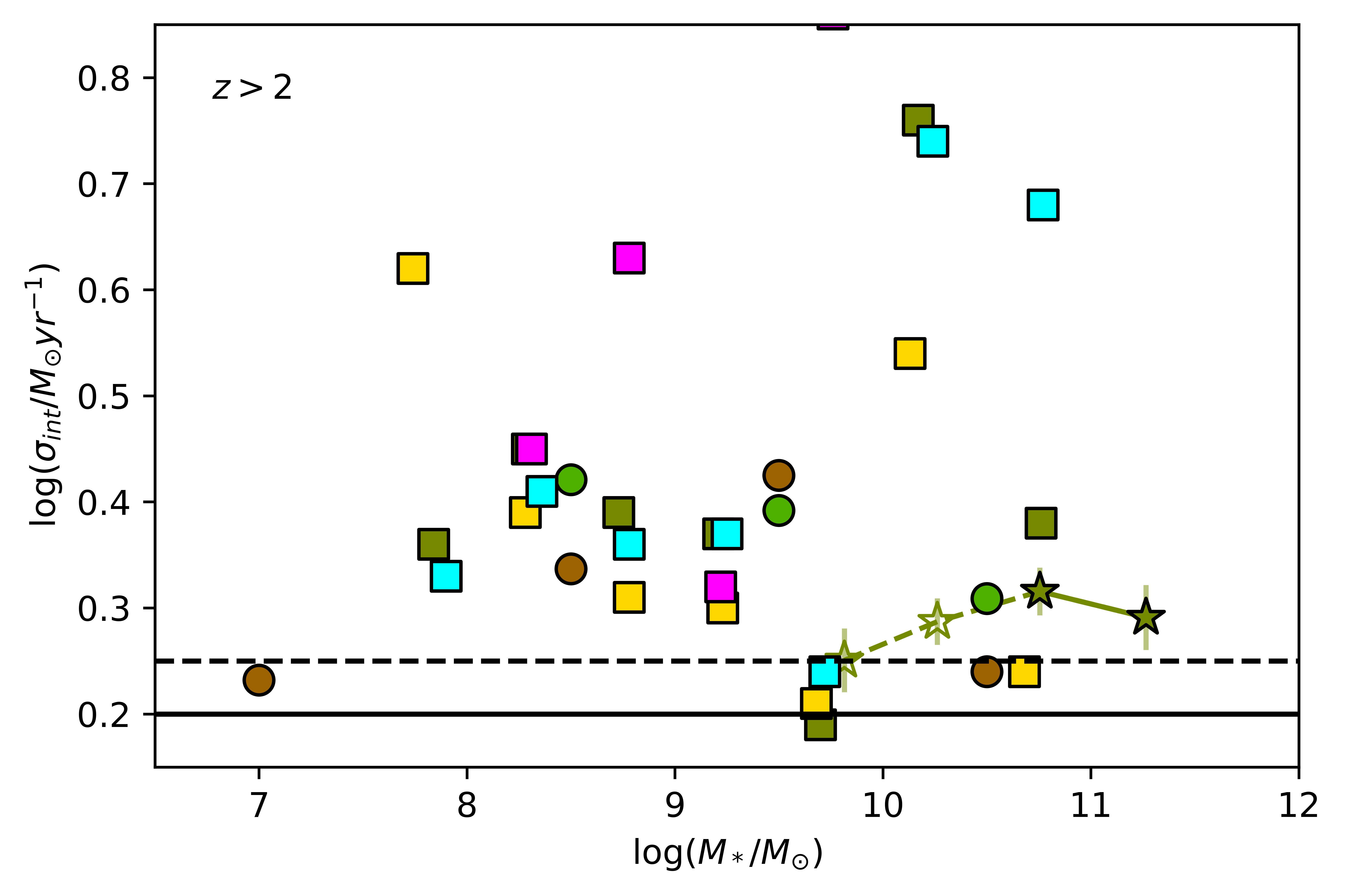} &
    \includegraphics[width=0.48\textwidth]{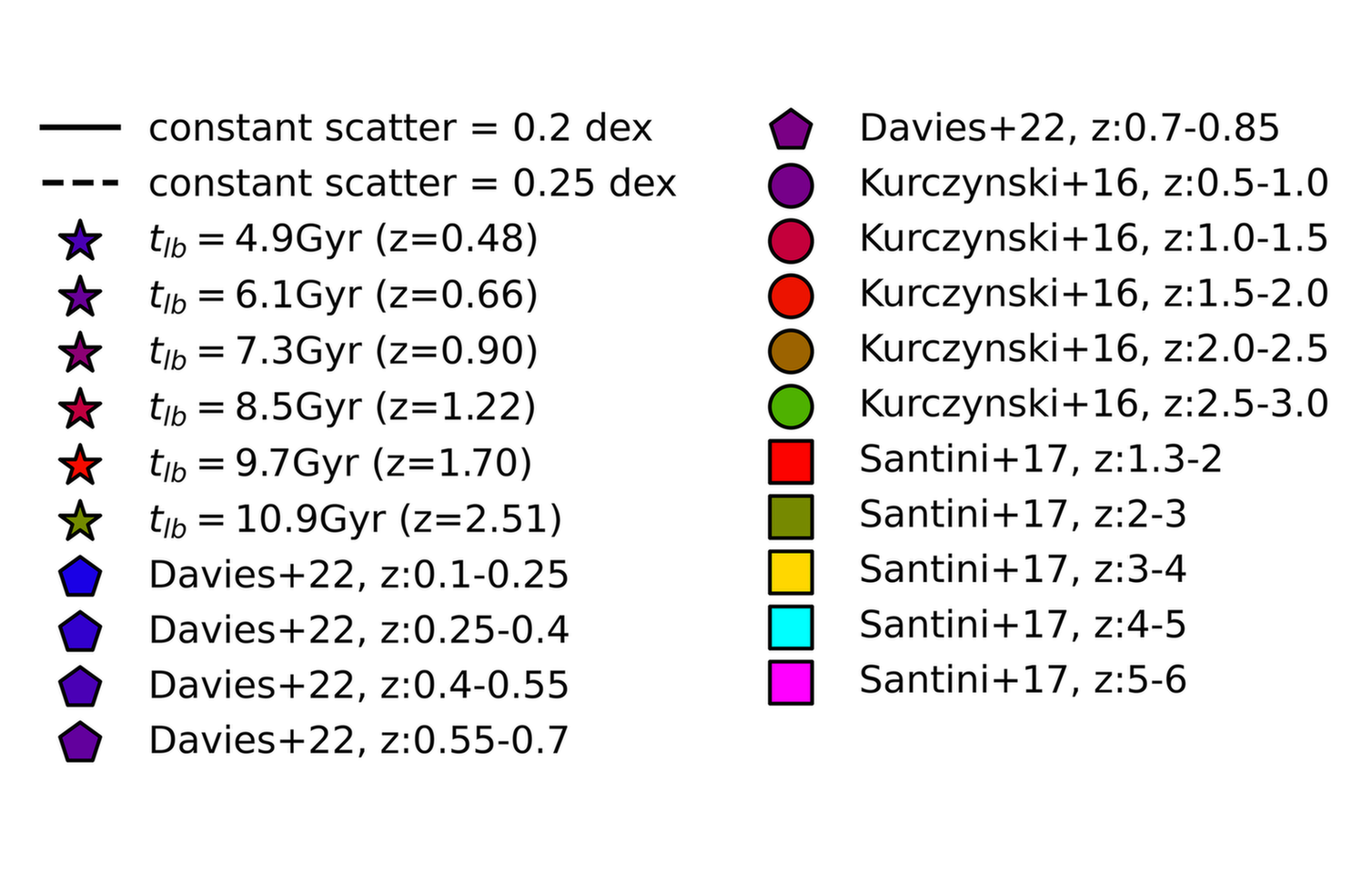} \\ 
    \end{array}
    $$
    \caption{A comparison of our intrinsic MS scatter to previous studies. The first three panels display the distribution of intrinsic MS scatter over $M_*$ at $z_{{\rm phot}} < 1$, $1 < z_{{\rm phot}} < 2$ and $z_{{\rm phot}} > 2$, respectively. The star symbols are the results in this study. The pentagon symbols show the results from \citet{davies22}, while the curves in the top left panel are polynomial fitting curves to their data. The filled pentagons are the  fitting range of \citet{davies22}. The circles are the results from \citet{kurczynski16}, while the squares represent the data from \citet{santini17}. The horizontal dashed line in the bottom indicates a constant intrinsic MS scatter at 0.25 dex \citep{daddi07, ciesla14, speagle14}, while the solid line represents a non-evolving scatter at 0.20 dex \citep{noeske07, whitaker12, pessa21}. Our data are qualitatively similar to \citet{kurczynski16, davies22}. }
    \label{fig: Comparision}
\end{figure*}

We compare our measurements of the intrinsic MS scatter with some previous studies in Figure \ref{fig: Comparision}. Traditionally, a redshift-independent width of MS at either 0.2 or 0.25 dex is founded in observations \citep{daddi07, noeske07, whitaker12, ciesla14, speagle14, kurczynski16}, which is lower than our results. In contrast, \citet{kurczynski16, santini17, tacchella20, davies22, shin22} report a redshift evolution of intrinsic MS scatter, which spans a wider range from 0.2 to 0.9 dex. 

Our data show a similar trend to the `U-shaped' distribution described in \citet{davies22}. They used the DEVILS survey, containing $\sim$60,000 galaxies with spectroscopic redshifts ranging from $0.1$ to $0.85$. In the case of $z_{{\rm phot}}=0.5$, where our samples have redshift overlap, the intrinsic MS scatter trend along the $M_*$ is highly consistent with the `U-shaped' distribution, except we have significantly smaller intrinsic MS scatter (0.24 - 0.40 dex in this study, while $\sim$0.4 - 0.8 dex in \citet{davies22}). At low stellar mass, \citet{davies22} suggest that the intrinsic MS scatter is driven by stochastic starbursts and stellar feedback events; while the galaxies become more massive and reach intermediate stellar mass (around $10^{10}M_{\odot}$), the galaxies are too massive so that the effect of star formation and galaxy feedback is less significant. In this study, intrinsic MS scatter increases at high stellar mass ($\log(M_*/M_{\odot}) \gtrsim 10.3$), consistent with the `U-shaped' distribution. \citet{davies22} conclude that AGN feedback leads to a large scatter at the high mass end. We note that in our selection, we removed sources with \textit{current} AGN signatures (see Section \ref{COSMOS Sample}), but that the feedback from \textit{previous} AGN will still affect the MS scatter over longer timescales than the AGN duty cycle. As the redshift increases, more data in the low stellar mass end are identified as mass-incomplete due to IR selection. However, we can still recognize that the intrinsic MS scatter tends to increase when galaxies become more massive for $M_*>10^{10}M_{\odot}$. With increasing redshift, we notice that the right half of the `U-shaped' distribution becomes flattered and even decreases at $t_{{\rm lb}} = 10.9$Gyr ($z_{{\rm phot}}=2.51$). This suggests that the star-forming and feedback activity or efficiency for high-mass galaxies in the early universe may differ relative to lower redshifts. 

Regarding the higher intrinsic MS scatter in \citet{davies22} relative to our results, we think this is due to the following reasons: (1) They do not include photo-z uncertainties in the SED modeling, which results in an underestimate of the measurement uncertainty (on $M_*$ and SFR), and hence, an overestimation of the intrinsic MS scatter. (2) They derive galaxy properties for the D10 field of DEVILS \citep{davies18} by using different SED fitting code, \texttt{PROSPECT} \citep{robotham20}. Large differences in obtained properties are produced by different derivation techniques applied to various photometric data (see the comparison between \texttt{PROSPECT} and \texttt{MAGPHYS}\footnote{We pick \texttt{MAGPHYS+photo-z} in this study because this code can treat redshift as a free parameter and derive the $z_{\rm phot}$ and the corresponding measurement uncertainty.} in \citet{thorne21}). (3) They adopt the U-V-J selection rather than sSFR cut so that the samples in \citet{davies22} contain low-sSFR galaxies (see the discussion of these two selection criteria in Appendix \ref{sSFR selection vs U-V-J Cut}). However, in our selection criteria, these galaxies are identified as quenched and removed. The addition of quenched galaxies severely enlarges the MS scatter. As will be shown in Section \ref{Comparison to theoretical studies} when comparing to simulations, the amplitude of the intrinsic MS scatter is very sensitive to the choice of sSFR cuts. For example, \citet{leja22} demonstrate that fixed-sSFR cuts may reduce the inferred MS scatter, particularly at the highest stellar mass. (4) They do not strictly require detection in the IR bands. As discussed in Section \ref{Contribution of including IR data to the Uncertainties of M_* and SFR}, the SED fitting in the absence of IR information results in considerable uncertainty of derived SFR, which may increase the SFR standard deviation and inferred intrinsic MS scatter.

\subsection{Comparison to theoretical studies}
\label{Comparison to theoretical studies}

\begin{figure*}
    \centering
    $$
    \begin{array}{cc}
    \centering
    \includegraphics[width=0.48\textwidth]{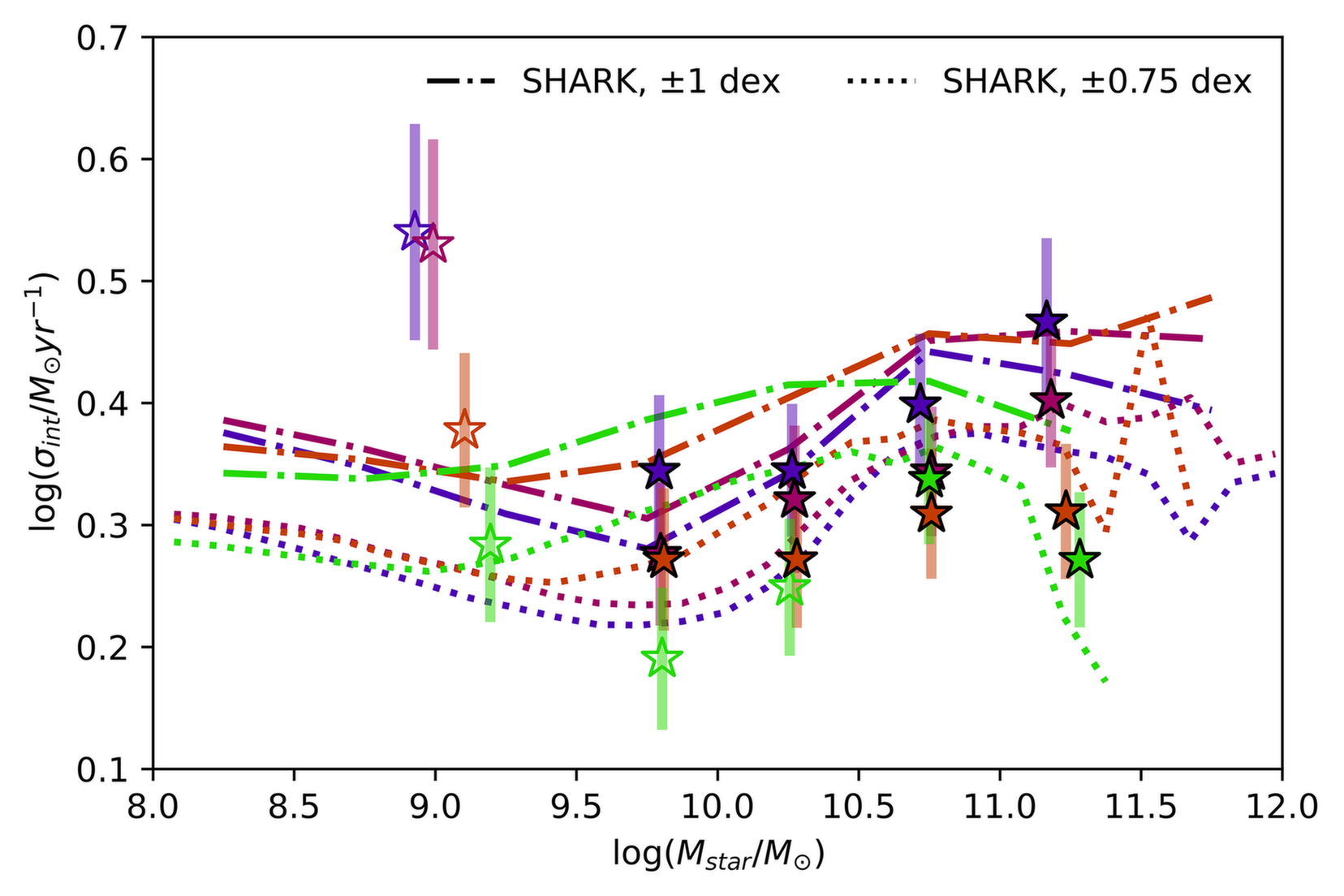} &
    \includegraphics[width=0.48\textwidth]{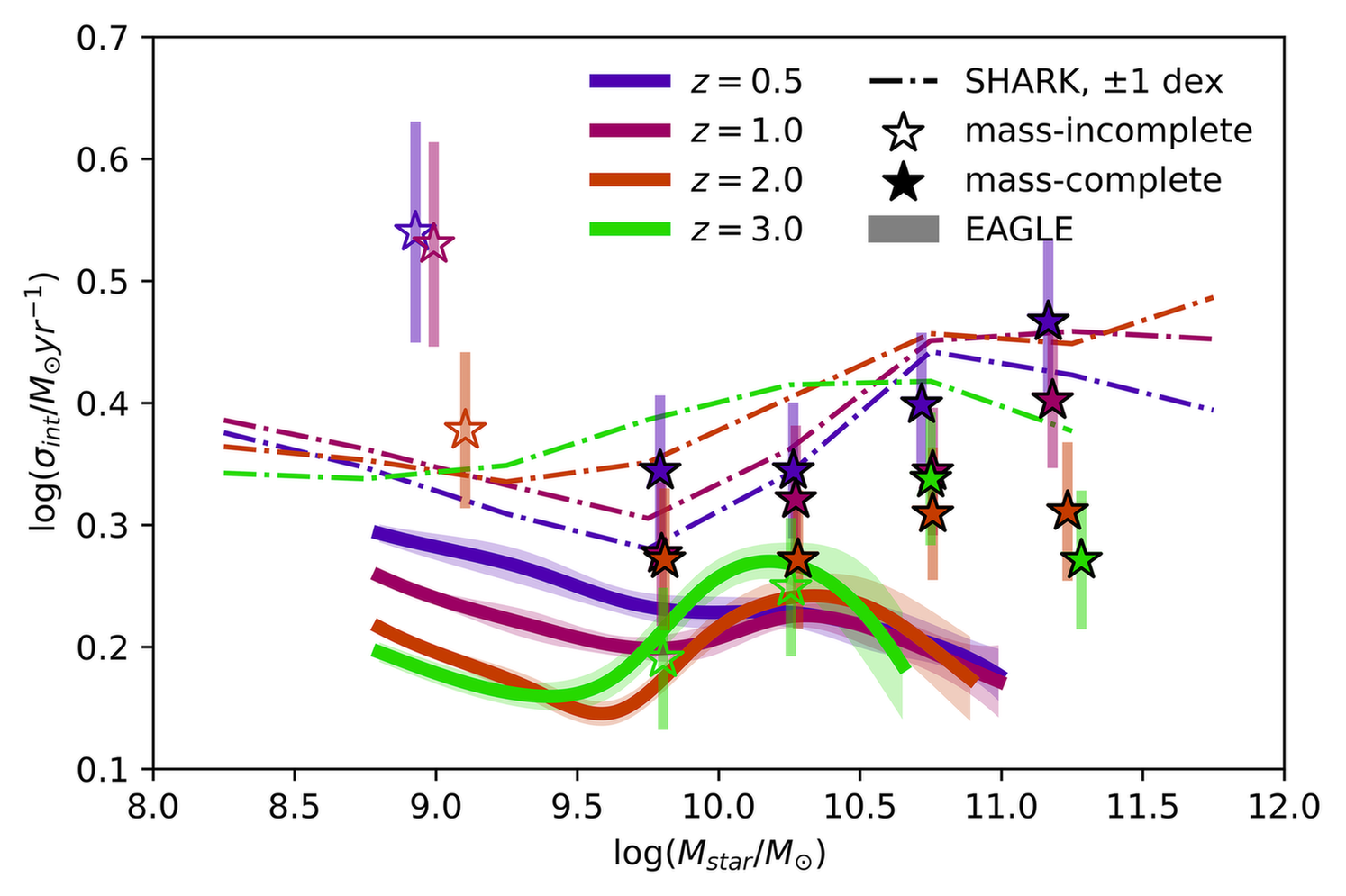}
    \end{array}
    $$
    \caption{Left panel: we compare our results with \texttt{SHARK} models and show the sensitivity of the intrinsic MS scatter to the chosen boundary of sSFR (i.e., $\pm1$ and $\pm0.75$ dex). The trends in \texttt{SHARK} show rough qualitative agreement with our observational results but with slight differences in the normalization, which is sensitive to the adopted sSFR cut. Right panel: we overlap the results from the \texttt{SHARK} (dot-dashed lines) and \texttt{EAGLE} (solid bands) simulations \citep{lagos18, Matthee19} with our data. For both panels, we have modified our bins to be consistent with the bins used in \citet{Matthee19}. At $M_*>10^{9.8}M_{\odot}$, where we are mostly mass-complete, we find differing trends between the observational results and the \texttt{EAGLE} simulations, and also note there are large differences in the trends inferred from \texttt{SHARK} and \texttt{EAGLE}. We obtain error bars of intrinsic MS scatter by bootstrap resampling the data by their uncertainties 100 times and remeasuring the intrinsic scatter. }
    \label{fig: simulation}
\end{figure*}

We first compare to results from the \texttt{SHARK} semi-analytic models \citep[left panel of Figure \ref{fig: simulation};][]{lagos18}. For \texttt{SHARK}, we select galaxies with stellar masses between $10^{8.25}-10^{11.75}M_{\odot}$ within $\pm1$ dex along the MS for each redshifts. We calculate the standard deviations of the median SFRs for selected galaxies and present the results as the dot-dashed lines in Figure \ref{fig: simulation}. A notable difference from the observational data is that the overall scatter in \texttt{SHARK} is larger than observational data in the mass-complete range, particularly for the highest redshift bin. We find a common trend that the \texttt{SHARK} results follow a similar `U-shaped' distribution at each redshift, though the minimum and maximum points occur at a stellar mass $<10^{10}M_{\odot}$ and $>10^{10.75}M_{\odot}$, respectively. We suggest that the consistent `upturn' feature in \texttt{SHARK} also indicates the effect of \textit{past} AGNs for $M_* \gtrsim 10^{10}M_{\odot}$. We also observe a flat or decreasing scatter in \texttt{SHARK} for galaxies in the range of $M_* \geq 10^{10.75}M_{\odot}$. This occurrence indicates that galaxies in this mass range are shifted below the chosen sSFR limit because AGNs have a more significant impact on them. We also investigate the effect of the sSFR cut on intrinsic MS scatter and find that the stricter sSFR cut leads to a smaller amplitude of the intrinsic MS scatter. For instance, the intrinsic MS scatter will be reduced by $\sim 1$ dex overall when we pick a selection with a narrower sSFR cut, such as $\pm 0.75$ dex along the MS, rather than $\pm1$ dex.  
 
Next, we compare the results with \texttt{EAGLE} hydrodynamical simulations \citep[right panel of Figure \ref{fig: simulation};][]{Matthee19}. We redivide the data in redshift binning at $z_{{\rm phot}} = 0.5$, 1.0, 2.0 and 3.0 for a proper comparison with Figure 3 in \citet{Matthee19}. \citet{Matthee19} adopt SF galaxies with evolved sSFR selection (i.e., $\log({\rm sSFR}/{\rm yr}^{-1}) = -10.4$ at $z = 0.5$ and increases to $-9.4$ at $z = 3$) and measure the scatter from the residuals by using the non-parametric local polynomial regression method. Then they obtain the intrinsic MS scatter by subtracting the observational errors derived by median uncertainties of the observational sample from \citet{chang15}. The \texttt{EAGLE} results suggest lower intrinsic MS scatter at higher redshift for $M_*<10^{9.8}M_{\odot}$ and similar intrinsic MS scatter for $M_*>10^{9.8}M_{\odot}$, with a downward `U-shaped' feature appearing at z = 2 and 3. This shape differs substantially from our findings. Unlike the decreasing trend with stellar mass from simulation, the intrinsic MS scatter in this study decreases initially but increases at the high mass end. \citet{Matthee19} consider that supermassive black hole growth accounts for the increasing scatter at $M_* \approx 10^{9.8}M_{\odot}$ at high redshift. However, our results suggest the influence of the feedback mechanism from \textit{previous} AGNs might be more significant at higher stellar masses ($M_* \gtrsim 10^{10.3}M_{\odot}$) at low redshift. The intrinsic MS scatter at each redshift bin is also larger than \citet{Matthee19}. We suspect that different physics (e.g., feedback, SF model) adopted in \texttt{EAGLE} give rise to the quantitative difference to our results and from \texttt{SHARK}.

\section{Conclusions}
\label{Conclusions}

In this paper, using the selection of 12,380 SF galaxies from the COSMOS2020 database and adopting the \texttt{MAGPHYS+photo-z} SED fitting code, we characterise the intrinsic scatter of galaxy MS over the redshift range $0.5<z<3.0$. 

\begin{itemize}
  \item We find that the intrinsic MS scatter is larger than the measurement uncertainty by a factor of 1.4-2.6 when IR data is available for accurately constraining the dust-obscured star formation (Section \ref{Results and Analysis}), with measured MS scatter in the range of 0.26-0.47 dex. 
  \item For the COSMOS2020 sample, the inclusion of IR data is the dominant factor (over $z_{{\rm phot}}$ uncertainty), affecting the accuracy of measuring the scatter on the MS. 
  \item Binning the data according to either redshift or look-back time, we find a slightly negative correlation between intrinsic MS scatter and look-back time (Equation \ref{eq: IS vs z_phot}) but with an upturn at $t_{{\rm lb}} \gtrsim 10$Gyr. 
  \item To connect the intrinsic MS scatter with the feedback mechanism, we present a toy model that uses the \citet{behroozi10} SMHM relation (Equation \ref{eq: IS = k_i*f_HMSM}), which does a reasonable job of matching the distribution of intrinsic MS scatter over $M_*$ and redshift (Figure \ref{fig: IS vs mass}), although with less agreement at the highest redshifts. 
  \item We compare our results to other observational studies of the MS scatter. Differing from a non-evolving, mass-independent scatter, our results are qualitatively similar to the `U-shaped' intrinsic MS scatter distribution with stellar mass and redshifts found in \citet{davies22}. 
  \item We compare the intrinsic MS scatter to some theoretical studies. The consistent upturn trend in \texttt{SHARK} models suggests the agreement of the feedback mechanism from past AGN activity for galaxies with $M_* > 10^{10}M_{\odot}$. These comparisons highlight the significant influence that sSFR cuts can have on the measured value of the `intrinsic' MS scatter and that particular care needs to be taken with such comparisons. We also find that the behaviour of intrinsic MS scatter diverges significantly between our study and \texttt{EAGLE} Simulation.  
\end{itemize}  

In the future, the most significant gain in our understanding of the evolution in the MS scatter will come from deeper surveys in rest-frame IR to enable accurate characterization of the MS scatter at both low stellar masses and higher redshifts, where our current sample is severely limited. There is a weak agreement between observation data and theory in Equation \ref{eq: IS = k_i*f_HMSM} for high-$z$ and low mass cases. In particular, better sampling in these regimes will provide a clear test of whether our toy model linking the MS scatter to the HMSM relation is reasonable. Alternatively, we also plan to explore less-restrictive selection criteria in the IR bands to push our sample to include more low-$M_*$ and/or high-$z$ sources from existing surveys.

 \section*{Acknowledgements}
 
The authors appreciate the referee, Antonios Katsianis, who provided valuable and insightful comments on the manuscript. Parts of this research were supported by the Australian Research Council Centre of Excellence for All Sky Astrophysics in 3 Dimensions (ASTRO 3D), through project number CE170100013. We thank the COSMOS team for making the data products that we used in this project. K.G. is supported by the Australian Research Council through the Discovery Early Career Researcher Award (DECRA) Fellowship DE220100766 funded by the Australian Government. We also thank Jorryt Matthee for correspondence regarding his study using the \texttt{EAGLE} survey. This research was conducted on Ngunnawal Indigenous land.

 \section*{Data Availability}
The observational data used in this paper are publicly available through catalog and imaging data releases from the COSMOS survey team (see Section \ref{COSMOS Sample}). Other data products can be made available upon reasonable request to the first author.



\bibliographystyle{mnras}
\bibliography{Rongjun.bib} 

\appendix
\label{appendix}
\section{\texorpdfstring{\MakeLowercase{s}{S}}SFR selection vs U-V-J Cut}
\label{sSFR selection vs U-V-J Cut}

\begin{figure*}
     \centering
    $$
    \begin{array}{cc}
    \centering
    \includegraphics[width=0.39\textwidth]{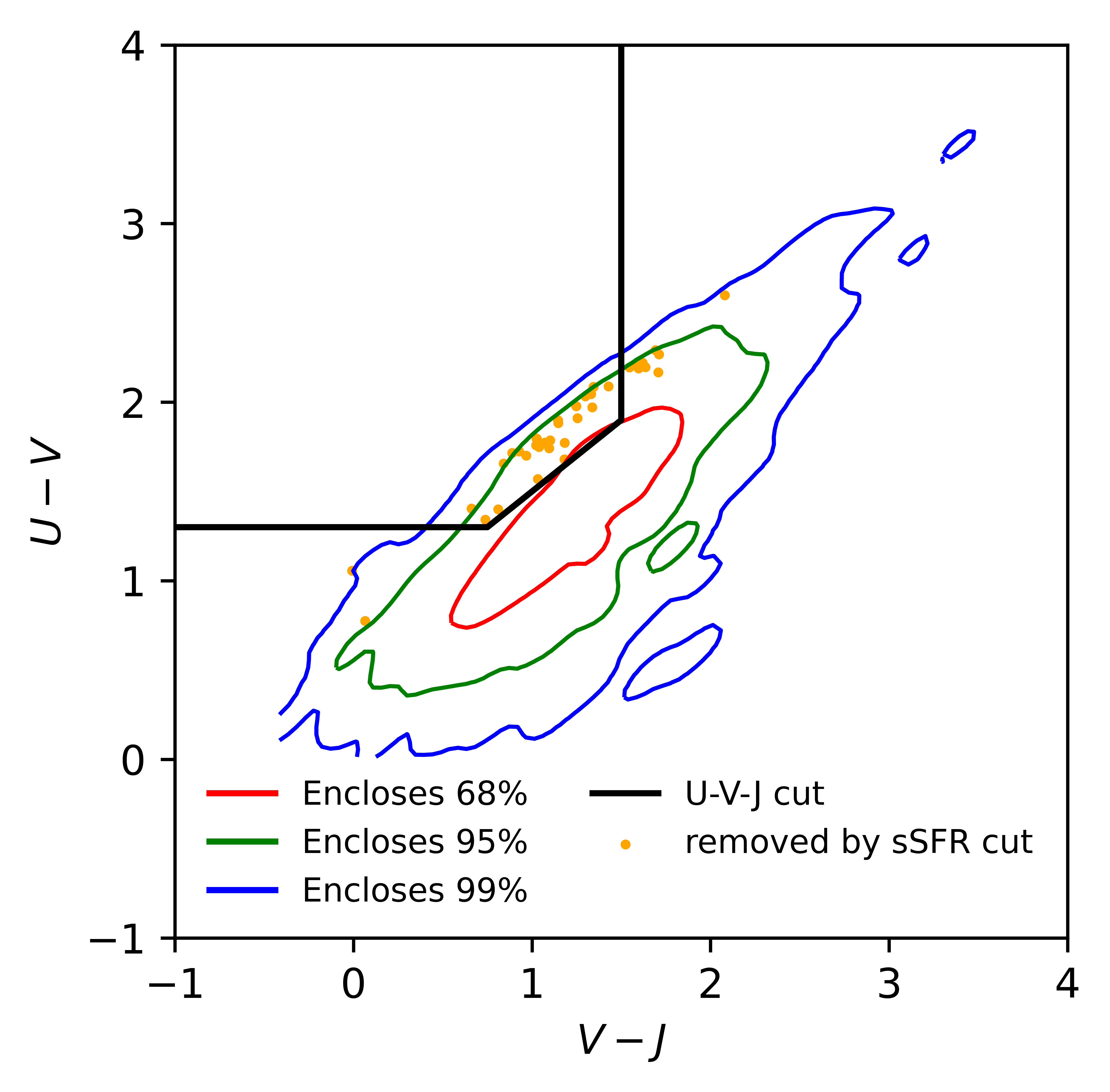} &
    \includegraphics[width=0.57\textwidth]{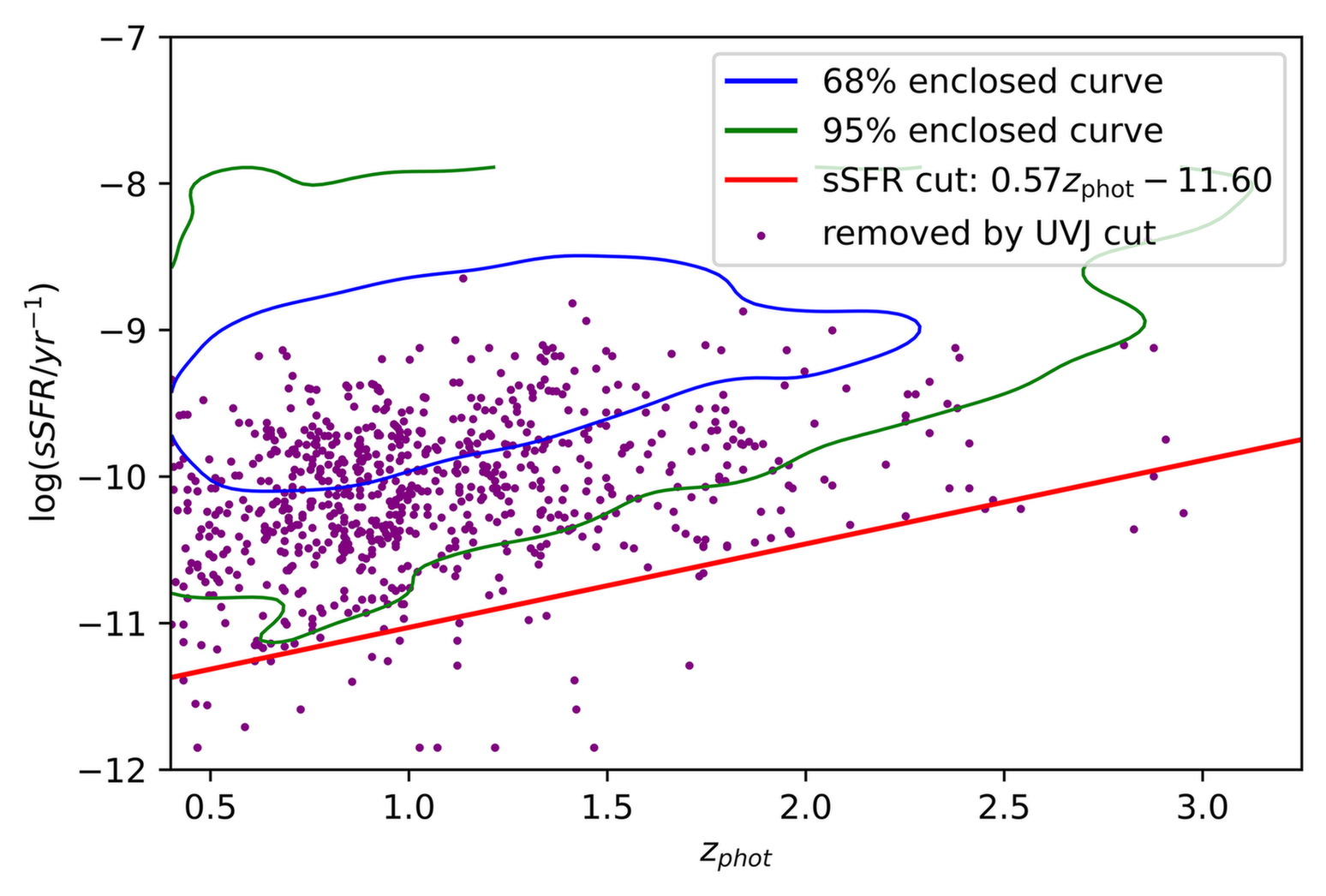}
    \end{array}
    $$
    \caption{Left panel: U-V-J diagram lot for the 13,071 $\chi^2$ selection galaxies. Right panel: in addition to Figure \ref{fig: sSFR cut}, we show the galaxies removed by U-V-J cut from \citet{whitaker12}. 64 galaxies are excluded by the sSFR cut, and 929 galaxies are excluded by the U-V-J cut, whereas only 48 of them are marked as quenched galaxies jointly by both selections. }       
    \label{fig: sSFR vs U-V-J}
\end{figure*}

We show a comparison between the conventional U-V-J (in rest-frame) selection from \citet{whitaker12} and our adopted sSFR selection (see Figure \ref{fig: sSFR vs U-V-J}). Since the sSFR is computed by multi-band SED fitting, we expect this technique will be more accurate in excluding quenched galaxies than a color-color cut based on U, V, and J bands. We find that $75\%$ of galaxies below our sSFR cut are excluded by the U-V-J cut, so the majority of `quenched' galaxies in our samples are also designated as `passive' in the U-V-J diagram. However, only $5.17\%$ galaxies that were eliminated by U-V-J selection lie below our sSFR cut. Hence, the bulk of galaxies excluded by the colour-colour cut for our sample are SF galaxies; presumably, they are dusty SF galaxies. Therefore, we adopt an sSFR cut rather than U-V-J cut to remove quenched galaxies for our analysis. 

\label{lastpage}
\end{document}